\documentclass[12pt]{article}

\usepackage[utf8]{inputenc}
\usepackage[margin=1in]{geometry}

\usepackage{amsmath,amsthm,amssymb,bbm}

\usepackage{graphicx}
\usepackage[table,xcdraw]{xcolor}
\usepackage{subfig}

\usepackage{booktabs}
\usepackage{longtable}
\usepackage{makecell}

\usepackage{enumerate}
\usepackage{enumitem}
\usepackage{caption}
\usepackage[normalem]{ulem}
\usepackage{setspace}
\usepackage{mdframed}
\usepackage{accents}

\usepackage[title]{appendix}
\usepackage{xr}

\usepackage{authblk}

\usepackage{natbib} 
\usepackage{hyperref}

\usepackage{pdfpages}

\usepackage{comment}

\newcommand{\blind}{1}

\setcitestyle{authoryear}

\hypersetup{
    colorlinks=true,
    citecolor=blue,
    linkcolor=blue,
    filecolor=blue,      
    urlcolor=magenta
    }

\newcommand{\Nf} {\mathcal{N} }
\newcommand{\R}{\mathbb{R}}

\DeclareMathOperator{\diag}{diag}

\newcommand{\by}{\textbf{y}}

\newcommand{\bz}{\textbf{z}}

\newcommand{\bw}{\textbf{w}}

\newcommand{\bZ}{\textbf{Z}}

\newcommand{\bbeta}{\boldsymbol{\beta}}

\newcommand{\btheta}{\boldsymbol{\theta}}

\newcommand{\bmu}{\boldsymbol{\mu}}

\newcommand{\bepsilon}{\boldsymbol{\epsilon}}
\newcommand{\boldeta}{\boldsymbol{\eta}}
\newcommand{\bLambda}{\boldsymbol{\Lambda}}
\newcommand{\blambda}{\boldsymbol{\lambda}}

\theoremstyle{definition}

\begin{document}

\def\spacingset#1{\renewcommand{\baselinestretch}%
{#1}\small\normalsize} \spacingset{1}


\if1\blind
{
\title{\bf Leveraging ontologies to predict biological activity of chemicals across genes}
\author[1]{Jennifer N. Kampe}
\author[1]{David B. Dunson}
\author[3]{Celeste K. Carberry}
\author[3]{Julia E. Rager}
\author[2]{Daniel Zilber}
\author[2,4]{Kyle P. Messier}

\affil[1]{Duke University, Department of Statistical Science}
\affil[2]{National Institute of Environmental Health Sciences, Division of Intramural Research, Biostatistics and Computational Biology Branch} 
\affil[3]{University of North Carolina at Chapel Hill, Gillings School of Global Public Health, Department of Environmental Sciences and Engineering} 
\affil[4]{National Institute of Environmental Health Sciences, Division of Translational Toxicology, Predictive Toxicology Branch} 
  \maketitle
} \fi

\if0\blind
{
  \bigskip
  \bigskip
  \bigskip
  \begin{center}
    {\LARGE\bf Leveraging ontologies to predict biological activity of chemicals across genes}
\end{center}
  \medskip
} \fi

\bigskip
\begin{abstract}
High-throughput screening (HTS) is useful for evaluating chemicals for potential human health risks. However, given the extraordinarily large number of genes, assay endpoints, and chemicals of interest, available data are sparse, with dose–response curves missing for the vast majority of chemical–gene pairs. Although gene ontologies characterize similarity among genes with respect to known cellular functions and biological pathways, the sensitivity of various pathways to environmental contaminants remains unclear. We propose a novel Dose–Activity Response Tracking (DART) approach to predict the biological activity of chemicals across genes using information on chemical structural properties and gene ontologies within a Bayesian factor model. Designed to provide toxicologists with a flexible tool applicable across diverse HTS assay platforms, DART reveals the latent processes driving dose–response behavior and predicts new activity profiles for chemical–gene pairs lacking experimental data. We demonstrate the performance of DART through simulation studies and an application to a vast new multi-experiment data set consisting of dose–response observations generated by the exposure of HepG2 cells to Per- and polyfluoroalkyl substances (PFAS), where it provides actionable guidance for chemical prioritization and inference on the structural and functional mechanisms underlying assay activation.

\end{abstract}

\noindent%
\vfill

\newpage
\spacingset{1.9} 

\section{Introduction}
\label{sec:dart_intro}

Since the mid-20th century, the global use of industrial chemicals has exploded, with more than 86,000 registered for commercial use in the U.S. alone. Approximately 62\% of the total chemical volume produced consists of substances classified as hazardous to human health \citep{unep2023}. Industrial activity releases large quantities of chemicals into the environment and incorporates both deliberate additives and accidental contaminants into many finished products. Consequently, humans are exposed to tens of thousands of industrial chemicals through air, water, food, consumer products including textiles, cosmetics, and toys, and occupational activities. Recent biomonitoring studies have detected chemicals of concern in human tissues and secretions, including heavy metals such as lead and cadmium, and persistent organic pollutants such as dioxins and PFAS in breast milk \citep{Serreau2024}, PFAS \citep{kotlarz2020measurement} and microplastics \citep{leonard2024microplastics} in blood, and bisphenol-A (BPA) \citep{Li2013} and triclosan \citep{Allmyr2008influence} in urine. The economic impact of chemical and heavy metal exposures, including illness, reduced cognitive function, healthcare costs, and lost productivity, was estimated to exceed 10\% of global GDP in 2017 \citep{Grandjean2017}. 

Per- and polyfluoroalkyl substances (PFAS), commonly known as ``forever chemicals,'' are a class of thousands of man-made compounds widely used in products such as nonstick cookware, firefighting foams, textiles, and food packaging due to their resistance to water, heat, and stains. Their strong carbon-fluorine bonds make them highly persistent, leading to accumulation in the environment, marine food chains, and the human body. Exposure to PFAS -- through food, drinking water and consumer products -- has been linked to a variety of adverse health outcomes, including immune system dysfunction \citep{ehrlich2023consideration}, decreased fertility \citep{bach2016perfluoroalkyl}, pre-eclampsia \citep{abd2025maternal}, developmental delays \citep{oh2021prenatal}, and kidney and testicular cancer \citep{steenland2021pfas}. 

Given the substantial human, environmental, and economic costs of chemical exposures, there is strong interest in identifying hazardous chemicals and establishing regulatory standards. Since 1976, the U.S. Environmental Protection Agency (EPA) has maintained the Toxic Substances Control Act (TSCA) Chemical Substances Inventory, which lists over 86,000 chemicals, approximately 42,000 of which are currently active in U.S. commerce, with new additions made regularly \citep{epaTSCA}. While new chemicals must demonstrate an appropriate safety profile, pre-existing chemicals are largely exempt unless the EPA demonstrates toxicity. Under the current framework, most chemical volume is subject only to voluntary industry testing \citep{Hall2012}. Nonetheless, advances in molecular biology and toxicology have enabled large-scale expansion of toxicity testing.

In 2007, the U.S. National Research Council articulated a vision for large-scale toxicological testing to match the scope of modern chemical exposure \citep{nrc2007tox}. Central to this vision was a shift from traditional animal-based tests to in vitro and high-throughput approaches. High-throughput screening (HTS) automates testing across numerous chemicals and gene assays, producing dose-response data to prioritize chemicals for further study. The EPA Toxicity Forecaster (ToxCast), introduced in 2007, was the first large-scale implementation of this strategy, evaluating thousands of chemicals across hundreds of assays \citep{richard2016toxcast}. Building on this, Toxicology in the 21st Century (Tox21) --  a collaboration between the EPA, NIH, and the FDA -- combines HTS with assay development and validation of new testing paradigms \citep{Thomas2018}. While these programs provide publicly available data for thousands of chemicals, coverage across gene assay targets is often sparse due to logistical constraints. Despite significant progress, only a small fraction of known chemicals has been adequately tested for toxicity \citep{plass2025estimating}, leaving an urgent need for scalable toxicological screening and modeling approaches.

\subsection{Data description and scientific challenges}
\begin{figure}[htp!]
    \centering
    \includegraphics[width=0.9\linewidth]{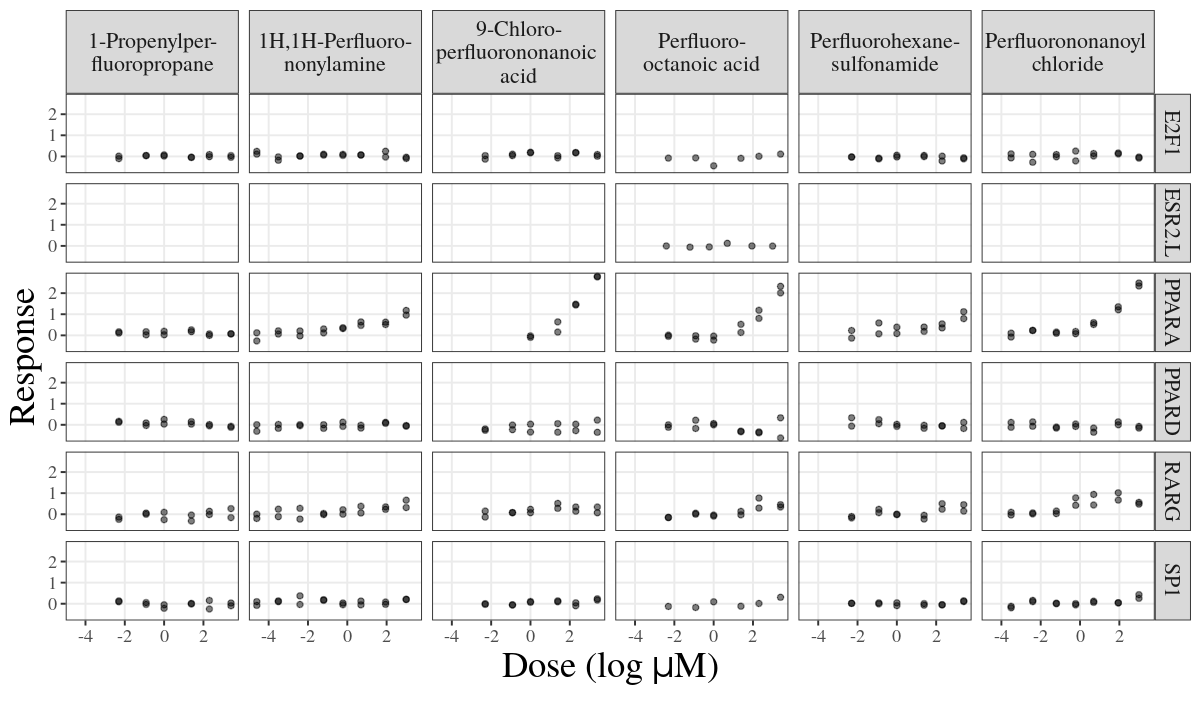}
    \caption{Observed dose-response data for a subset of the data including six gene-specific assay endpoints (rows) and six PFAS chemicals (columns). Each cell contains the log$_2$ fold-induction responses elicited at each tested dose (natural log scale) for the respective pairs, with some pairs lacking data entirely.}
    \label{fig:EDA_sample_curves}
\end{figure}

To address the need for extensive toxicological screening, this work aims to (1) elucidate modes of action for biologically active chemicals and (2) predict responses for untested assay–chemical pairs in sparse HTS data. Typical HTS datasets prioritize broad chemical coverage, with tens of thousands of chemicals in EPA databases \citep{kavlock2012, kleinstreuer2017development}, but are limited to a narrow set of assay endpoints—often only hundreds representing a small subset of biologically relevant genes \citep{huang2014profiling}. Combining datasets across studies introduces further challenges, such as irregular dose grids due to varying concentration protocols. These issues are illustrated by the PFAS-HepG2 dataset, which includes (1) legacy HTS data from the NIH Integrated Chemical Environment with 123 assay endpoints and 204 PFAS chemicals tested on HepG2 cells, and (2) a new dataset measuring responses of three PFAS chemicals across 10,645 assay endpoints. The data matrix, indexed by assay endpoints (rows) and chemicals (columns), is dominated by inactive or missing dose–response curves, many of which are sparse or truncated (Figure \ref{fig:EDA_sample_curves}). While the combined data allow borrowing of information across assay-dense and chemical-dense submatrices, existing approaches struggle with such extremely sparse, matrix-structured functional data.

\subsection{Background}

Quantitative Structure–Activity Relationship (QSAR) modeling is central to virtual screening, prioritizing chemicals for further study based on structural features associated with biological activity. In the single assay context, \citet{Wheeler2019} proposes a Bayesian Additive Adaptive Basis Tensor Product (BAABTP) model, which uses Mold2 QSAR descriptors to learn basis functions for the dose-response surface. In practice, this often imposes excessive similarity among structurally similar chemicals, even when structure is poorly aligned with activity. This issue is addressed by decomposing chemical latent factors into structure-only and structure-activity components \citep{Moran}. While these methods leverage structure to predict dose–response curves, they are limited to single-assay settings.
integration, is crucial for robust inference.

Pharmacological approaches that match tumor cell lines to effective drugs parallel efforts in environmental health to identify active chemical–gene pairs. Despite large-scale automated testing, methods lag in denoising curves and imputing untested combinations. Multilevel factor models improve drug sensitivity estimates by borrowing strength across drugs and cell lines \citep{vis2016multilevel}, and have been extended to functional matrix completion via Bayesian tensor filtering \citep{Tansey2022}. Multi-output Gaussian processes integrate chemical and genomic covariates for joint curve prediction \citep{Gutierrez2024}, but all assume dense, regular dose grids—unsuited to sparse, irregular HTS data.

In toxicology, matrix completion methods are designed for sparse, irregular assays. \citet{Wilson} use a zero-inflated spline model with assay random effects, though fixed chemical effects proved too rigid. \citet{Jin2020} model both biological activity and heteroskedasticity using shared latent factors and spline-based curves, enabling flexible dose–response estimation. While addressing multiple challenges, their approach assumes relatively dense data and omits chemical covariates—highlighting the need for stronger cross-chemical and cross-gene information sharing in the sparse HTS setting.

Gene co-regulation creates dependencies across assays, offering opportunities for structured information sharing. \cite{Melo2023} show that gene networks are highly complex with simple assortative methods failing to capture connectivity patters. Functional grouping has mostly been used in post-processing \citep{Davidson2022}, but offers clear potential for borrowing information across genes. However, joint modeling approaches must accommodate substantial dependence and heterogeneity: assays vary in sensitivity, and the corresponding genes may be co-regulated \citep{Wilson}. 

In latent factor GP models, functional grouping typically targets the covariance structure, which may be too restrictive for heterogeneous co-regulated genes. Recent work by \cite{Schiavon} proposes a new class of factorization models which instead target the sparsity structure of the loadings matrix. This approach induces dependence through meta-covariate-informed shrinkage, allowing sparsity patterns to vary across factors.

Recent work highlights the potential of flexible, covariate-informed models for dose-response inference, but most methods assume regular dose grids and denser observations than are typical in environmental health. Additionally, covariate-informed methods are largely limited to the single assay, multiple chemical case in environmental health applications. In response, we propose a novel Gaussian Process model for dose-response curve estimation that incorporates gene- and chemical-level covariates through hierarchical shrinkage priors on gene-specific latent Gaussian processes and latent factor regression on chemical features. Our approach builds on the infinite factorization priors of \cite{Schiavon}, modeling genes as conditionally independent given meta-covariates while inducing dependence through shared sparsity patterns. This framework enables principled information sharing across genes and chemicals and supports robust inference in the highly sparse, irregular settings characteristic of modern environmental screening studies.

\section{Data}

 The proposed model is motivated by the PFAS-HepG2 dataset, which combines (1) a new set of experiments exposing three PFAS chemicals to a large collection of HepG2 liver cell assay endpoints (hereafter UNC24), and (2) all PFAS–HepG2 experiments available from the NIH Integrated Chemical Environment as of May 6, 2024 (hereafter ICE). Details on the individual datasets and the harmonized combined dataset are provided below.

\subsection{UNC24 PFAS mixtures data}

The UNC24 PFAS mixtures dataset \citep{Carberry2025} consists of experiments exposing HepG2 immortalized liver cells to three PFAS—Perfluorooctanoic acid (PFOA), Perfluorooctanesulfonic acid (PFOS), and Perfluorohexanoic acid (PFHxA)—at concentrations of 2.5, 10, 25, and 35~$\mu$M across six experimental replicates. PFOA and PFOS are widely used to make products resistant to sticking, heat, water, grease, and fire, while PFHxA is a common breakdown byproduct of other PFAS. Gene expression was measured for 10,645 probe sets, each corresponding to a collection of DNA probes targeting a single gene, representing 10,247 distinct genes in total.  Raw expression values are normalized across the dataset using Signal Space Transformation–Robust Multi-chip Analysis (SST-RMA) in the Affymetrix Transcriptome Analysis Console. The response is recorded as the $\log_2$-transformed SST-RMA fold change, $\log_2\frac{Y_1}{Y_2}$.

\subsection{HepG2 PFAS ICE data}

While the UNC24 dataset provides a complete matrix of gene expression responses measured on a regular grid of doses, we are additionally motivated by a larger and more irregular dataset spanning a broader set of chemicals. This second dataset includes all experiments involving PFAS chemicals and the HepG2 cell line from ICE \citep{EPAice}. Chemicals were drawn from the EPA’s \href{https://comptox.epa.gov/dashboard/chemical-lists/PFASMASTERLISTV2}{PFASMASTERLISTV2} list \citep{EPApfaslist}, which currently includes 7,993 PFAS compounds. Unlike the UNC24 data, which follows a consistent protocol across all experiments, the ICE data are highly heterogeneous with respect to dose and replicate protocols. 

The ICE data report transcriptomic assay endpoints rather than gene expression probe sets. Gene-relevant endpoints, primarily from Attagene and Tox21 platforms, span 77 unique genes, 28 of which are not present in UNC24. Many of these additional genes are related to estrogen and androgen receptor activity, making them of particular interest for evaluating potential endocrine disruption by PFAS. After filtering to remove non-gene-related assays and incompatible response formats (see Supplement), the cleaned ICE data comprises 177,118 observations across 64 single-gene assays and 193 PFAS chemicals.

\subsection{Combined data set} \label{data:combined}

The combined HepG2–PFAS dataset comprises 815,818 experiments across 10,269 genes and 193 chemicals, evaluated at up to 40 dose levels ranging from 0.01 to 300~$\mu$M. Because most biological activity occurs below 25~$\mu$M and high-dose observations are sparse, we log-transform doses to increase resolution at lower concentrations and discretize the transformed scale for computational efficiency.

\begin{figure}[htp!]
    \centering
    \includegraphics[width=0.85\linewidth]{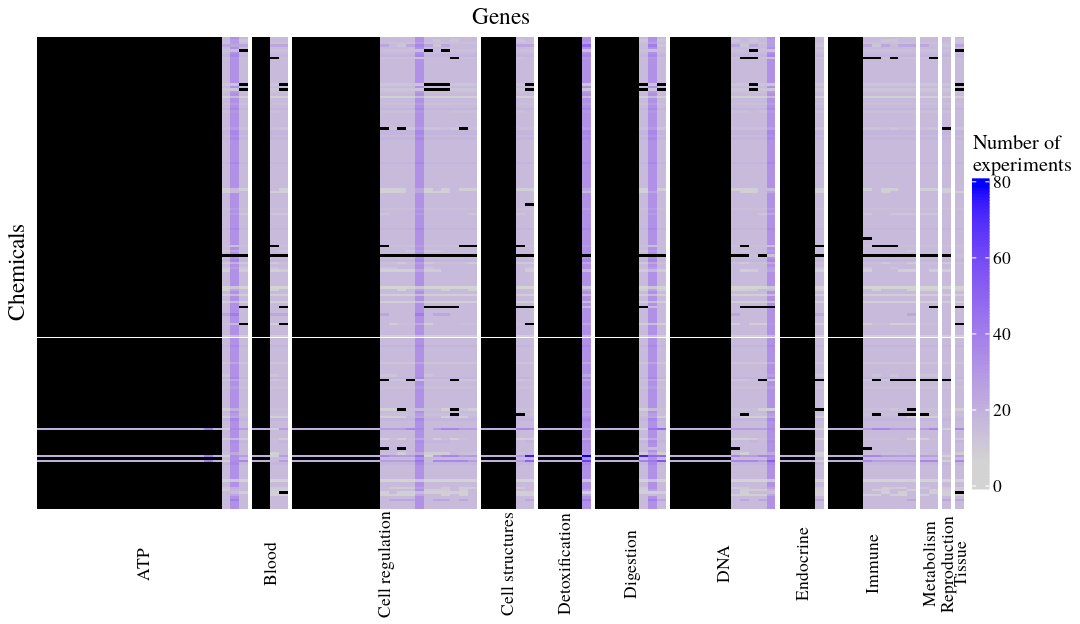}
    \caption{Overview of the combined PFAS-HepG2 data showing the distribution of experiments among the 193 included PFAS chemicals and a subset of 100 genes. Cell color indicates the number of experiments performed with each pair, with each dose counted as a separate experiment; untested pairs are indicated by a black cell. Genes are organized by functional annotation, with genes associated with multiple biological pathways grouped with the most common pathway to which they are annotated.}
    \label{fig:EDA_heatmap}
\end{figure}

The dataset is highly sparse, with 97\% of gene–chemical pairs unobserved. Two dense submatrices provide opportunities for information sharing: one includes 10,206 genes exposed to the same three PFAS chemicals, and the other contains 55 genes tested against nearly all PFAS chemicals. Figure~\ref{fig:EDA_heatmap} illustrates this structure, with UNC24 observations covering many genes and few chemicals, and ICE observations covering many chemicals and few genes.

Replicate variability is generally low, with rare instability or batch effects, consistent with the controlled conditions typical of HTS using immortalized cell lines. However, we observe dose-dependent heteroskedasticity, with greater between-replicate variance at higher doses, and gene-varying heteroskedasticity. 

Consistent with prior findings \citep{Wilson, Jin2020}, certain genes show widespread activation across chemicals, whereas chemical effects vary more strongly across genes. Genes that are activated across chemicals are likely related to pathways such as xenobiotic metabolism and represent a response to foreign molecules rather than a specific chemical response. Regardless of the cause for elevated baseline response, appropriate models must be able to capture this heterogeneity across genes. 

\subsection{Gene ontology covariates} \label{sec:dart_GO}

Gene ontology covariates are derived from the Hallmark Gene Sets \citep{liberzon2015molecular}, a curated collection of gene sets based on replicated associations in the literature. Of the 10,269 genes in the dataset, 2,834 have at least one annotation, though these annotations are sparse (median prevalence $<1$\%). To reduce dimensionality and enhance interpretability, we collapse the 50 Hallmark sets into 12 broader pathways and represent each gene with binary indicators denoting pathway membership. Because genes may contribute to pathways in yet undiscovered ways, these covariates are treated as presence-only.

\subsection{Chemical covariates} \label{sec:dart_qsars}

Chemical covariates are derived from QSAR models of physicochemical properties using the \href{https://github.com/kmansouri/OPERA}{OPERA software} \citep{Mansouri2024QSAR}. These include structural, physicochemical, and ADME-related descriptors, with only point estimates retained. The resulting 32 real-valued covariates are reduced via principal component analysis to the first five components, explaining 76.7\% of the variance.

\vspace{0.5cm}

Motivated by the block structure of the data and the availability of high-quality covariates, our model leverages dense submatrices to infer dose–response functions for sparsely observed pairs, while accounting for dose-dependent heteroskedasticity. To support imputation for sparsely observed chemicals and genes, we incorporate structural chemical covariates and functional gene covariates, reflecting the premise that structurally similar chemicals and functionally related genes exhibit similar activity. Because related genes may be measured on assay platforms with varying sensitivity and noise, the framework must flexibly adapt to differences in measurement quality across endpoints.

\section{Covariate-informed latent space model} 
\label{sec:dart_methods}

In the chemical--gene activity problem, we aim to model dose--response curves for a large number of gene assays indexed by  $j = 1, \ldots, M$, each measured under exposure to a subset of chemicals spanned by $i = 1, \ldots, N$. Observations are made on an irregular grid of dose levels $d = d_1, \ldots, d_D$ and across a variable number of experimental replicates $r = 1, \ldots, R$. To address the challenges of data sparsity and heterogeneity, we propose the \emph{Dose-Activity Response Tracking (DART)} framework---combining a multi-gene, multi-chemical generalized infinite factorization structure with dose- and gene-dependent heteroskedasticity. Specifically, the response for gene $j$ exposed to chemical $i$ in replicate $r$ is modeled as:

\begingroup
\setstretch{1}
\begin{align}
   \by_{ijr} &= \bmu_j + \bLambda_j \boldeta_i + \bepsilon_{ijr} \label{eq:lik} \\
   \bepsilon_{ijr} &\stackrel{iid}{\sim} \Nf(0, \Sigma_{j}), \quad \Sigma_{j} = \diag(\sigma^2_{1j}, \cdots, \sigma^2_{Dj}) \label{eq:hetero_cov} \\
   \log\sigma^2_{dj} &= \alpha_D + \gamma_j + \beta^D(d_d - \overline{d}) \quad \text{for } d = 1, \cdots, D  \label{eq:gene_dd_hetero}
\end{align}
\endgroup
where $\bmu_j \in \R^D$ is a gene-specific baseline dose-response curve, $\bLambda_j \in \R^{D \times K}$ is a collection of gene-specific latent Gaussian processes (GPs), and $\boldeta_i \in \R^K$ is a vector of chemical latent factors. The latent factors can be conceptualized as latent pathways through which chemicals may activate or suppress a gene's expression, with the corresponding GP's capturing the gene's susceptibility to each pathway across dose levels. Dose-level variation is incorporated into the gene latent components and the gene-level intercept based on the observed pattern of heterogeneous assay sensitivity as discussed in Section \ref{data:combined}. We model the heteroskedasticity in $\bepsilon_{ijr}$ under a log-linear variance model in Equation \eqref{eq:gene_dd_hetero} using the priors $\alpha_D \sim \Nf(0, \sigma_{\alpha_D})$, $\beta_D \sim \Nf(0, \sigma_{\beta_D})$. The gene-specific noise terms $\gamma_j$ in Equation~\eqref{eq:gene_dd_hetero} are modeled hierarchically as
$\gamma_j \sim \mathcal{N}(0, \tau_\gamma^2)$,
where the global scale parameter $\tau_\gamma$ controls the degree of heterogeneity across genes and is assigned a half-normal prior to encourage shrinkage toward homogeneity. This contrasts with the fixed-scale priors we place on $\alpha_D$ and $\beta_D$, which are easier to calibrate directly based on the observed range of dose-dependent variance in the data. In those cases, we set the prior scale conservatively to reflect heteroskedasticity observed in the data while avoiding explosive noise at high doses. For gene-specific noise, however, the appropriate level of heterogeneity is less clear \emph{a priori} and may vary widely depending on the subset considered, so we allow the data to inform it through the global scale parameter $\tau_\gamma$.

Replicate-level data are often averaged prior to modeling, discarding potentially important information about variability in the data generating process. In retaining replicates and treating them as noisy observations of some underlying mean effect $S_{ijd} = \bmu_j + \bLambda_j \boldeta_i$, we can improve uncertainty quantification and better separate systematic structure from measurement noise.  However, modeling replicate noise is not of primary interest as these fluctuations are not believed to be biologically meaningful or analytically relevant and so inference on the mean effect tensor $\mathcal{S} \in \R^{D\times N \times M}$ is of primary interest. 

\subsection{Baseline model: DART-NC}
While our primary model incorporates chemical and gene covariates to guide latent representations, we first introduce a flexible baseline version of the model that does not rely on any external covariates - Dose Activity Response Tracking with No Covariates (DART-NC). This formulation is especially useful in settings where chemical structural covariates or gene ontologies are missing or incomplete, and remains a valuable tool even when such data are available. In large and information-rich datasets, the latent factors alone may sufficiently capture the underlying signal, allowing the model to identify shared patterns of chemical activity and gene response directly from the observed data. 

To complete this model, we place independent priors on the chemical latent factors $\eta_i$ and gene-specific latent dose-response curves $\Lambda_j$. Specifically, we assume:

\begingroup
\setstretch{1}
\begin{align}
\boldeta_i &\sim \mathcal{N}(\mathbf{0}, \sigma_\eta^2 \mathbf{I}_K), \\
\blambda_{jk} \mid \tau_0, \gamma_k, C_k &\sim \mathcal{GP}(0, \tau_0 \gamma_k \, C_k), \ 
c_k(d,d') = e^{-(d-d')^2 / 2 l_c^2}, \\
\tau_0^{-1} &\sim \text{Ga}(g_\tau/2, g_\tau/2), \quad
\gamma_k \sim \text{MGP},
\end{align}
\endgroup
where length scale $l_c$ is given a uniform prior over a discrete grid, with reasonable values to include being informed by the range of the data, and MGP refers to the Multiplicative Gamma Process \citep{Bhattacharya.etal:2011}. Gene sensitivity is governed by global and column-specific shrinkage parameters $\tau_0$ and $\gamma_k$ respectively. These priors encourage sparsity and smoothness in the latent representations while allowing flexibility to capture gene-specific structure.

Because selection of the dimension $K$ for the latent space is significant for model performance yet difficult to determine a priori, we take the over-fitted factor model approach of \cite{Bhattacharya.etal:2011} and choose $K$ as a conservative upper bound and adaptively shrink non-informative dimensions. An appropriate value for $K$ should be determined on the basis of the complexity of the data in exploratory analysis, including the number of genes and chemicals modeled. Hence, the prior on $\gamma_k$ is specified as: 

\begin{equation*}
\label{mgp:prior}
\gamma_k = \prod_{s=1}^k \delta_s^{-1}, \quad 
\delta_1 \sim \mathrm{Gamma}(\alpha_1, \beta_1), \quad 
\delta_s \sim \mathrm{Gamma}(\alpha_2, \beta_2) \text{ for } s > 1
\end{equation*}

\subsection{Full covariate model: DART}

While the baseline model can flexibly recover structure in settings with sufficient data, it treats genes and chemicals as a priori exchangeable within their respective groups and does not leverage available biological or structural information. This limits its ability to generalize when data are sparse or unevenly observed. In particular, some chemical-gene pairs may be observed at only a few doses or not at all, making it challenging to infer dose–response patterns without additional regularization. To improve generalization and interpretability in such settings, we extend the model to incorporate auxiliary covariates for both chemicals and genes. 

Because the motivating dataset includes both well-studied and poorly characterized PFAS chemicals, we incorporate chemical structural covariates to encourage chemicals with similar physicochemical properties to share similar latent pathway representations, $\boldeta$. Let $\bw_i = (w_{i1}, \cdots, w_{iP})^\top$ be a vector of observed chemical structural covariates for each chemical $i$. Dependence between similar chemicals is induced via the latent factor regression, building on the approach of \cite{Moran}: 
\begin{equation}
w_{il} = \mu^c_{l} + \boldsymbol{\theta}_l^\top \boldsymbol{\eta}_i + e_{il}, \quad 
e_{il} \stackrel{\mathrm{iid}}{\sim} \Nf(0, 1), \quad 
i = 1,\ldots,N,\; l = 1,\ldots,P. 
\label{eq:eq_w}
\end{equation}

 We specify weakly informative independent normal priors for the chemical latent factors, the intercepts, and the factor coefficients in the covariate submodel: $\boldeta_i \sim \Nf_K(0, I)$, for $i = 1, \cdots, N$, $\mu^c_l \sim  \Nf(0, \sigma^2_w)$ for $l = 1, \cdots, P$, and $\btheta_l= (\theta_{l1}, \cdots, \theta_{lk})^\top \sim  \Nf_K(0, \sigma^2_\theta I)$ for $l = 1, \cdots, P$.
 
A persistent challenge in multi-assay studies is flexibly inducing dependence between the dose-response curves for related genes. Given recent work on gene co-expression networks \citep{Melo2023}, one intuitive approach is to introduce a gene assay variance-covariance matrix $C^g$, having entries $c^g(j,j') = d(j,j')$ where $d$ is some distance or similarity function defined on the genome. However, there is no broadly accepted measure of gene similarity, and we would like to avoid $\mathcal{O}(m^3)$ cost operations of the gene latent process covariance matrix. Another readily available source of prior information on the linkages between genes comes in the form of gene ontologies (i.e., gene sets or gene pathways) as discussed in Section \ref{sec:dart_GO}. Gene ontologies provide a vector of binary features for each gene which are believed a priori to be biologically relevant. To take advantage of these data, we propose a modification of the generalized infinite factorization prior of \cite{Schiavon} which models genes as being conditionally independent given meta-covariates and induces dependence via a shared sparsity structure. Let $\bLambda_j \in \R^{D \times K}$ be the matrix containing the latent dose-response GP's for gene $j$ as above in Equation \ref{eq:lik},  and let $\blambda_{jk}$ be its columns (each a dose-response GP). Additionally, let $z_{jl}$ be an indicator function for the association of gene $j$ with particular observable ontological pathways $l = 1, \cdots, Q$,  and let $\bz_j = (z_{j1}, \cdots,z_{jQ})^\top$. We specify the following local-global shrinkage prior for each latent GP $k$ and for each gene $j$:

\begingroup
\setstretch{1} 
\begin{align}
    \blambda_{jk} \mid \vartheta_{jk}, C_k &\sim \mathcal{GP}(0, \vartheta_{jk} C_k), \quad  c_k(d, d') = e^{-(d - d')^2/2l_c^{2}} \\
    \vartheta_{jk} &= \tau_0 \gamma_k \varphi_{jk} \\
        \tau_0^{-1} &\sim \text{Ga}(g_\tau/2, g_\tau/2) \\
        \gamma_k &\sim \text{MGP}\\
    \varphi_{jk}  &\sim  \text{LogNormal}(\mu_\varphi, \sigma_\varphi^2), \quad \mu_\varphi = \bz_j^T \bbeta_k -\sigma^2_\varphi / 2,  \label{eq:eq_DART_L}
\end{align}
\endgroup

\begin{figure}[!htp]
    \centering
\includegraphics[width=0.75\linewidth]{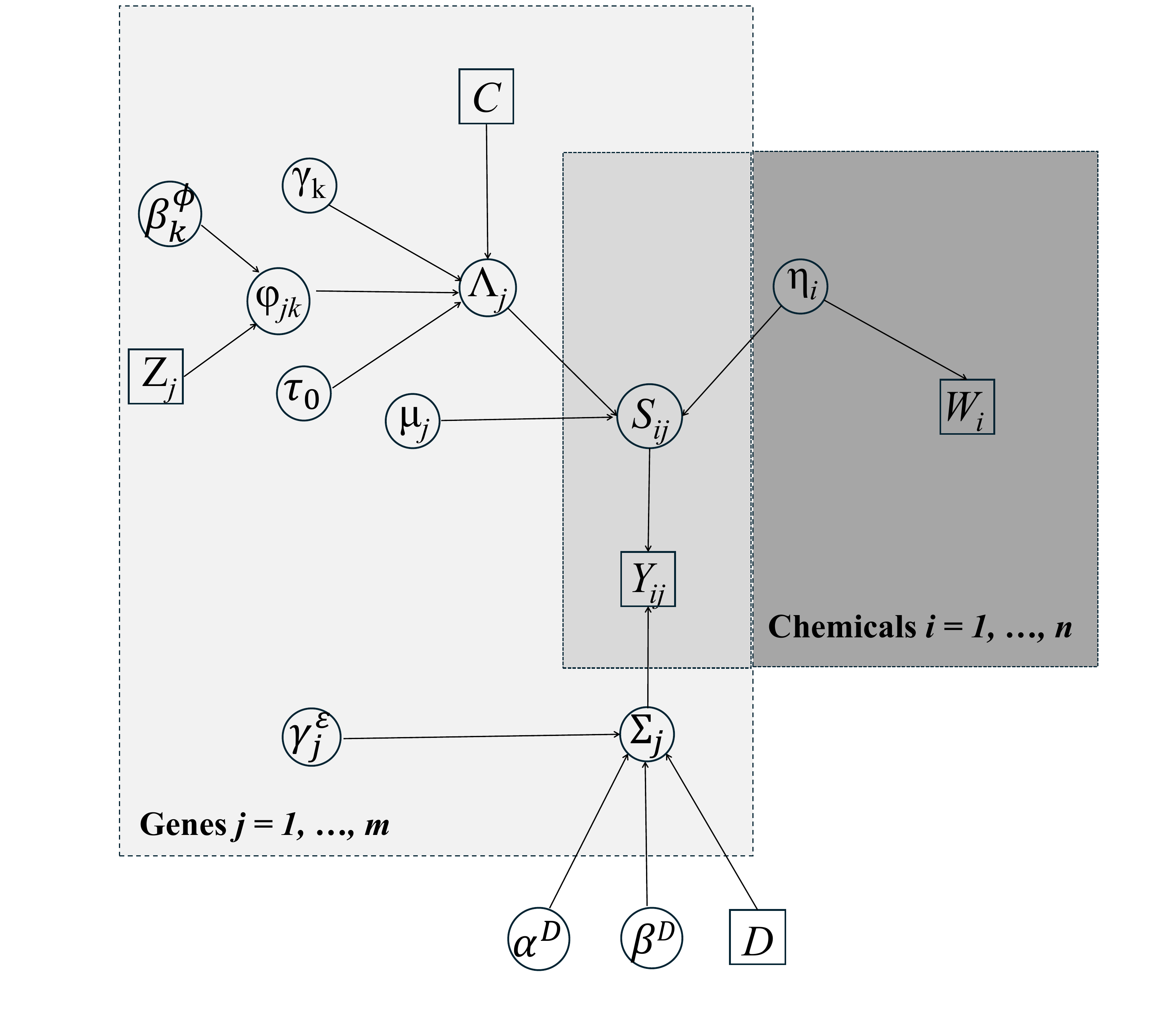}
 \caption{Directed acyclic graph (DAG) for the DART model. Squares are observed, circles latent; plates show replication over genes $j$ and chemicals $i$. Dose indexing ($d$) is omitted for clarity. Global terms ($C,\tau_0,\gamma_k$) are shown inside the gene plate for readability but do not vary with $j$.}
\label{fig:DAG}
\end{figure}

As in the baseline DART-NC model sparsity in the latent dose-response GPs is governed globally by $\tau_0$ and column-wise by $\gamma_k$; in addition, it is now governed locally by the covariate-informed term $\varphi_{jk}$---inducing dependence among genes with similar ontologies. As an example, consider two genes $j$ and  $j'$ with similar ontologies $\bz_j$ and $\bz_{j'}$. In this case, the model encourages similar variance parameters $\varphi_{jk}$, $\varphi_{j'k}$ across latent components, implying a comparable propensity for large responses - whether stimulating or inhibitory. Because this information borrowing operates through the GP variance rather than the mean function, genes with similar ontologies may still exhibit very different outcomes, including flat (inactive) or strongly active dose–response curves. 

An important consideration in this context is the handling of missing covariates. Gene ontology covariates are effectively presence only: zeros can indicate that a gene is not present in a given ontology or that the gene simply was not tested for the relevant pathway. In the proposed specification, we model $\varphi_{jk}$ with a log-normal prior with location parameter $\mu_\varphi = \bz_j^T \bbeta_k -\sigma^2_\varphi / 2$ such that $E[\varphi_{jk} \mid  \bz_j^T \bbeta_k = 0] = 1$. Hence, any missing covariates recorded as zero simply contribute to the shrinkage as multiplication by one is non-informative; if all covariates are recorded as zero, the entire local shrinkage coefficient $\varphi_{jk} = 1$ and so there is no local shrinkage. We choose $\sigma^2_\varphi$ to be small in consideration of the heavy tails of the log-normal. Additionally, given the sparsity of the gene ontology matrix $\bZ$,  we utilize a relatively informative normal prior centered at zero for the ontology coefficients $\bbeta_k = (\beta_{k1}, \cdots, \beta_{kQ})^\top$ to discourage overfitting with respect to rare non-zero gene ontology covariates: $\beta_{kq} \sim N(0, \sigma_0^2)$ independently for latent factors $k = l, \cdots , K$ and ontologies $q = 1, \cdots, Q$.

As in the baseline DART-NC model, selection of $K$ is handled via a multiplicative gamma process. An additional benefit in this extended model relates to a documented issue in the joint factor modeling literature \citep{Hahn01092013, palmer2025targetedempiricalbayessupervised}: when selecting $K$ using the joint likelihood of covariates and responses, high-dimensional predictors can dominate, leading to underestimation of latent factors that are important for modeling $y_i \mid \mathbf{x}_i$ but explain little of the total joint variation. In our setting, this risk is mitigated by the structure of $\beta_{kq}$: each latent pathway $k$ has its own associations with GO groups, allowing $\varphi_{jk}$ to counteract global shrinkage for genes in relevant pathways. While sparse, high-index pathways without GO annotations may still be missed, such cases tend to produce large residuals, signaling the potential for discovering novel groupings.

Because all of the parameters in Equations \eqref{eq:lik}-\eqref{eq:eq_DART_L} are continuous, we obtain posterior samples using No-U-Turn Hamiltonian Markov chain Monte Carlo \citep{hoffman2014nuts} efficiently implemented using within chain parallelization in the Stan probabilistic programming language. We demonstrate the ability of the model to recover true dose-response curves under conditions of varying sparsity through a simulation study, presented in the Supplementary Materials; we find that performance is highly robust to different levels of missingness --- even in the challenging cross validation scenario of pair-level holdouts -- for moderately sized networks. 

\section{Application to the PFAS-HepG2 data networks}
\label{sec:dart_app}

\subsection{Comparison to benchmark models}

We now turn to the primary scientific aim of the present study: using the proposed DART framework to impute missing dose-response curves from the PFAS-HepG2 HTS data. To assess predictive accuracy and benchmark performance, we begin with a focused comparison on a small, densely observed subset, where standard parametric models can be reliably fit. This setting provides a controlled environment to evaluate our model against established competitors before scaling to larger, more sparsely observed datasets.

 While many existing dose-response models for high-throughput screening (HTS) data either support matrix completion across multiple genes and chemicals or offer robust software, few provide both. The EPA’s widely used \texttt{tcplfit2} package offers well-maintained software but is limited to independent curve fitting for single gene-chemical pairs, with no support for missing data or prediction. To provide a realistic benchmark, we adapt \texttt{tcplfit2} to handle missing data and generate predictions, applying its most established models to a small, dense 10-by-10 gene-chemical subset, restricted to pairs with at least four distinct dose levels observed. Model performance is assessed via 5-fold cross validation, where fold $k$ holds out the $k$‑th lowest dose for each chemical–gene pair, ensuring at least four doses remain for model training.

In selecting competitor models from the \texttt{tcplfit2} package \citep{tcplfit2}, we utilize models that can accommodate various response curve shapes commonly observed in dose-response experiments: the Hill, Exponential 5 (Exp5), and Power functions \citep{judson2010invitro, thomas2019blueprint}. The Hill model is widely used, and is the default function in the US EPA's ToxCast Pipeline (TCPL) and parameterizes a sigmoidal dose-response via the E$_{max}$, EC$_{50}$, and slope. The power model fits the response as a three parameter power function of concentration, and while somewhat rigid, is useful for low-dose extrapolation and simple monotonic functions. The more flexible Exp-5 models the response as a four parameter exponential decay function in concentration, but as with the Hill and Power models, cannot capture non-monotonic responses. Additional details on these functions are provided in the Supplementary Materials.

\begin{figure}[htbp]
  \centering
  \includegraphics[width=0.85\textwidth]{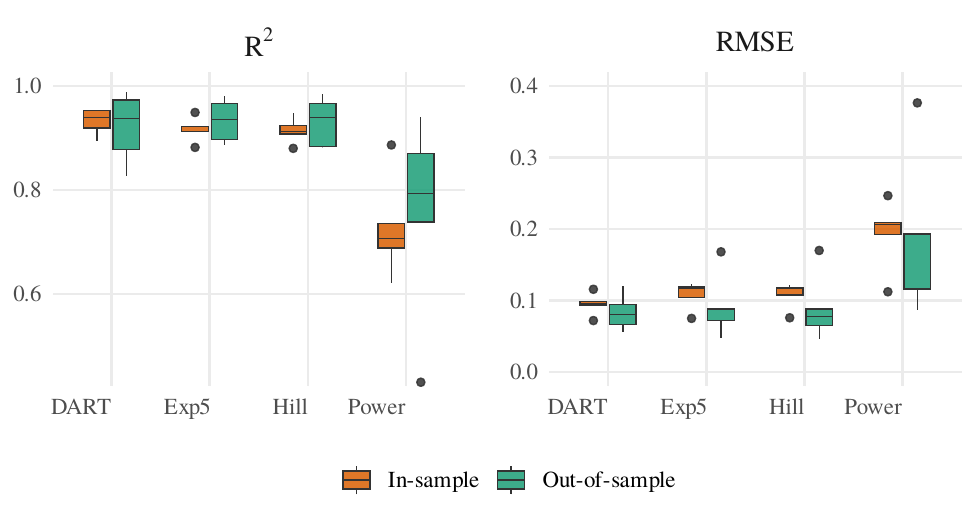}
  \caption{In-sample and out-of-sample performance (RMSE, $R^2$) by model across cross-validation folds, excluding large outlier poor performance folds for the Power model.}
  \label{fig:perf_boxplot}
\end{figure}

The DART model is fit via a single chain of 10,000 iterations while the TCPL models are run until reaching the specified convergence criteria. Figure~\ref{fig:perf_boxplot} summarizes in- and out-of-sample performance across cross validation folds. The proposed joint Bayesian DART model, fit to all gene--chemical pairs simultaneously with a modest latent dimension ($K = 10$), performs comparably to standard parametric dose-response models fit independently.

The DART model achieves similar accuracy to Hill and Exp5 in both $R^2$ and RMSE, while the Power model consistently underperforms; extremely poor performance for the Power model in fold 5 is excluded to improve visualization of the main trends. Notably, the Bayesian framework yields more stable predictions across folds, reflected in tighter distributions of in- and out-of-sample RMSE and the absence of outliers across folds, owing to information sharing via latent factors and covariates. Whereas $R^2$ follows the expected pattern of superior in-sample fit versus out-of-sample fit, RMSE is somewhat higher in-sample, a potential consequence of the cross validation scheme, which holds out one single dose per pair in each fold; because RMSE is unnormalized it is more sensitive to large errors in high response regions. In general, cross-validation errors are largest when intermediate doses are held out as these doses fall in the steepest part of the curve where small misspecifications in slope or curvature have the biggest effect; the Power model, which lacks an explicit asymptote parameter, provides an exception in that it performs worst when high doses are held-out. Additional details on per-fold performance are provided in the Supplementary Materials.

\begin{figure}[htbp]
  \centering
  \includegraphics[width=0.98\textwidth]{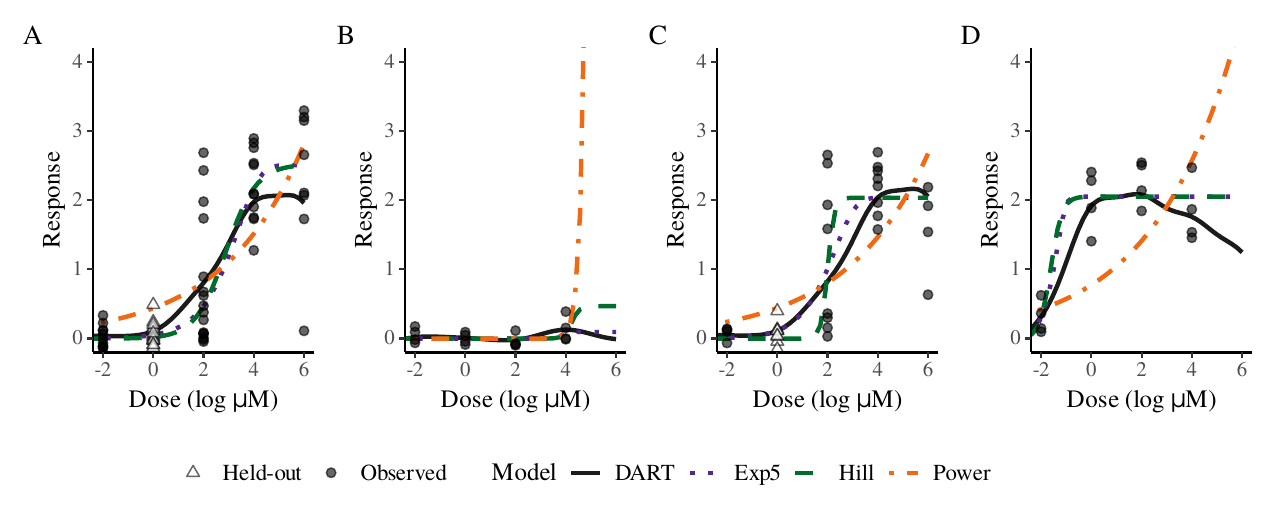}
  \caption{
    Predicted dose–response curves for selected chemical–gene pairs within one fold of cross validation. 
    Solid lines represent posterior means from the DART model with 95\% credible bands for the mean effect; 
    dashed and dotted lines represent benchmark models, points are observed dose-response combinations. Doses are displayed on the natural log scale.
    Panels show: 
    (A) PFOA-ESR1, 
    (B) (Heptafluorobutanoyl)pivaloylmethane-NR3C1, 
    (C) (Perfluoro-5-methylhexyl)ethyl 2-methylprop-2-enoate - ESR1, 
    (D) (Heptafluorobutanoyl)pivaloylmethane–NR1I2. 
  }
  \label{fig:tcpl_curves}
\end{figure}

Figure~\ref{fig:tcpl_curves} illustrates fitted curves for a representative subset of gene--chemical pairs from a single cross validation fold. Curves C and D lack sufficient dose levels to fit the TCPL  models with holdouts in this fold, and hence are fit on all points. As expected, DART, Hill, and Exp5 produce similar fits on monotonic responses, while the parametric models are unable to capture non-monotonic trends (e.g., Figure~\ref{fig:tcpl_curves}C). The DART model is specifically designed for sparse, structured data: while parametric models may offer slight advantages in dense, well-observed settings, DART uniquely supports joint inference and prediction under missingness---key capabilities for large-scale datasets such as ToxCast and ICE, which we investigate in the following experiment.

\subsection{Imputation in the PFAS-HepG2 Dataset}

We now turn to the motivating application: imputing unobserved portions of the dose–response surface for the PFAS–HepG2 dataset. The analysis focuses on a subset of 100 gene assays—including both annotated and unannotated genes—and all 193 retained PFAS chemicals. Across the data, 56\% of chemical–gene–dose triples are missing, and 45\% of chemical–gene pairs  are missing experimental data entirely, reflecting the sparsity typical of high–throughput toxicology screens.

To establish the reliability of inference and prediction, we first evaluated model fit through posterior predictive diagnostics and then quantified predictive accuracy using standard Bayesian information criteria and scoring rules. Goodness–of–fit was assessed using posterior predictive checks and coverage diagnostics, while out–of–sample predictive accuracy was measured using the widely applicable information criterion (WAIC), leave–one–out cross–validation (LOO), and the continuous rank probability score (CRPS), a proper scoring rule for full predictive distributions. We also conducted $k$–fold cross–validation, holding out all observations for $1/k$ of chemical–gene pairs in each fold. Models were fit to the full dataset using four chains of 25,000 MCMC iterations, and to each cross–validation fold using a single chain of 25,000 iterations. Full experimental details are provided in the Supplementary Materials.

 \begin{figure}
     \centering
     \includegraphics[width=0.75\linewidth]{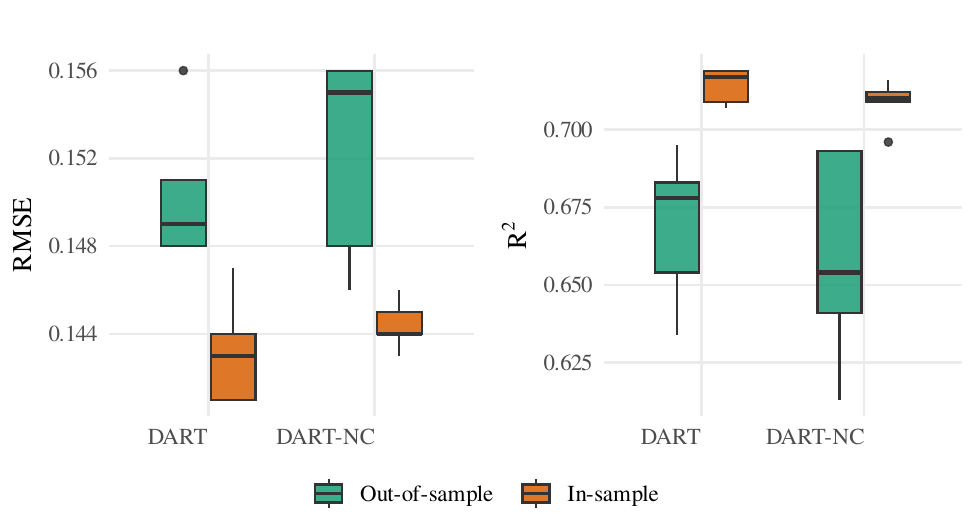}
     \caption{Performance of the DART and no-covariate baseline model, DART-NC across five-fold pair-structured cross validation.}
     \label{fig:CV_boxplot}
 \end{figure}

MCMC diagnostics indicated good convergence ($\hat{R} \leq 1.01$, effective sample sizes $ > 1000$). Posterior predictive checks show that the model captures the overall distribution of responses well, with slight oversmoothing near baseline activity. On average 95\% predictive intervals cover observations very close to the nominal level (92.5\%); coverage slightly exceeds nominal levels for low to moderate responses and drops below only for the highest response quintile. We observed a small group of points with very high empirical means around 2–4 on the log$_2$ scale (4x–16x induction relative to baseline) falling just outside their 95\% intervals. See Supplementary Materials Section 4.1 for additional details on MCMC and goodness of fit diagnostics. 

Next, we compare predictive performance across model variants using out-of-sample criteria (LOO, WAIC) and continuous rank probability score (CRPS), as well as $k$-fold cross-validation. In the full-data experiment, both models perform well, with the covariate model (DART) showing a slight advantage over the baseline (DART-NC). WAIC and LOO modestly favor the covariate model---$25.1 \pm 61.1$ (WAIC) and $45.8 \pm 62.2$ (LOO)---but the large standard errors make this evidence inconclusive. Mean CRPS was nearly identical, again giving a small edge to the covariate model and indicating similar predictive calibration.
Cross validation confirms the slight advantage of the covariate model in this application (see Figure \ref{fig:CV_boxplot}). Averaging across cross validation folds, DART achieves in- and out-of-sample $R^2$ of 0.714 and 0.669, respectively, compared to 0.709 and 0.659 for the DART-NC. Similarly, DART achieves in- and out-of-sample RMSE  of 0.143 and 0.150, respectively, compared to 0.144 and 0.152 for DART-NC.

\begin{figure}
    \centering
\includegraphics[width=0.95\linewidth]{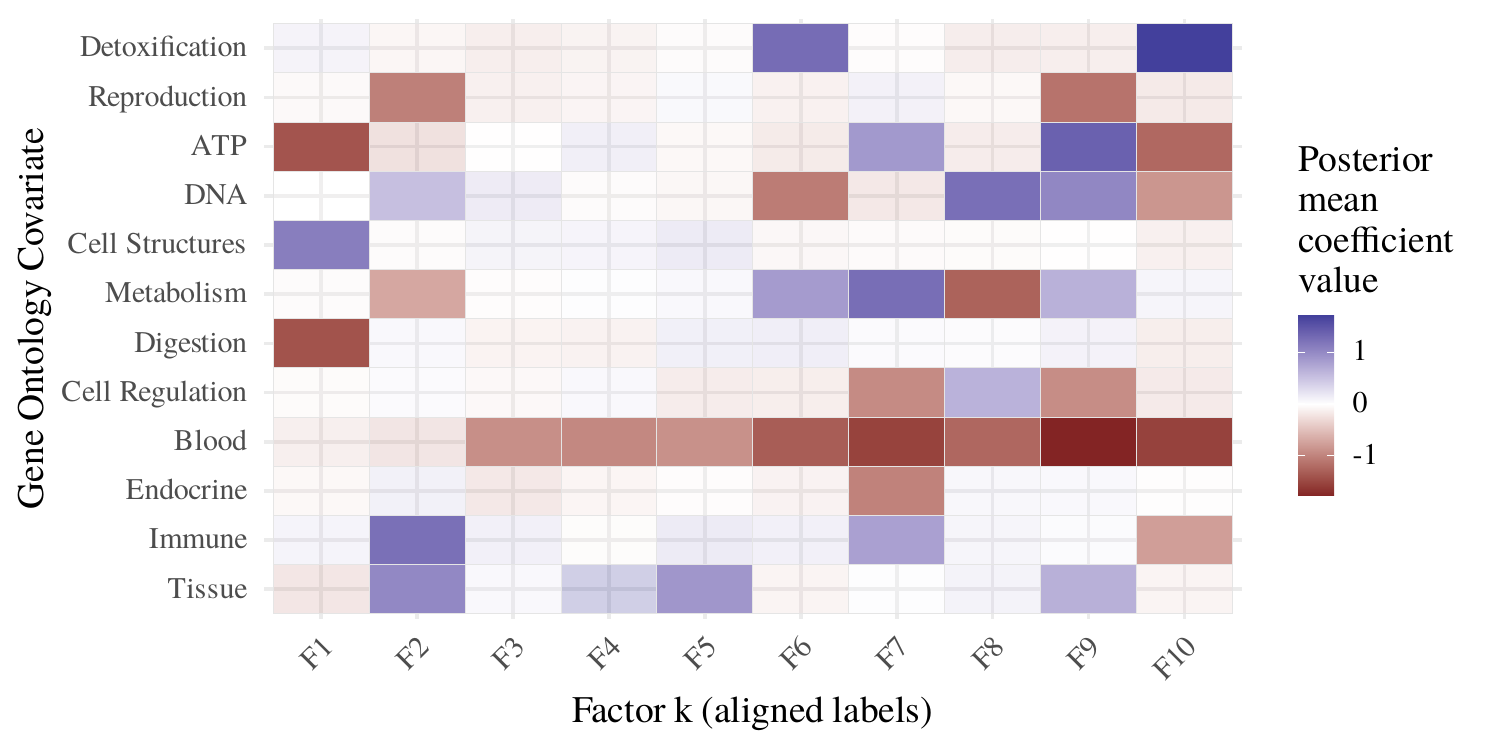}
    \caption{Gene-ontology covariate effects on aligned latent factors: color indicates posterior mean $\beta_{qk}$; opaque cells denote 95\% credible intervals excluding zero. Positive coefficients correspond to greater potential activity (either up- or down-regulation), while negative coefficients indicate stronger shrinkage toward flat responses.}
\label{fig:GO_covariates}
\end{figure}

The latent space alone captures the key features of dose–response behavior. Adding covariates yields further resolution, with several gene and chemical covariates showing significant associations that point to potential activity pathways. To resolve rotational ambiguity and label switching across posterior draws, we aligned the latent factor matrices using the MatchAlign procedure \citep{Poworoznek2021MatchAlign}, which applies per-draw Varimax orthogonalization followed by greedy matching for signs and labels, as implemented in the \texttt{R} package \texttt{infinitefactor} \citep{R-infinitefactor}. Figure~\ref{fig:GO_covariates} shows posterior mean estimates for gene coefficients whose 95\% credible intervals exclude zero. Under the log-normal prior on $\varphi_{jk}$, positive $\beta_{qk}$ values increase the expected $\phi_{jk}$ for genes in gene set $q$, reducing shrinkage and allowing greater flexibility in latent dimension $k$; negative $\beta_{qk}$ imply greater shrinkage toward null response. Negative effects dominate in the blood and reproduction gene sets, while positive effects dominate in detoxification, metabolism, immune, and tissue pathways. Among the chemical covariates, the influence of the chemical structural covariates (QSARS) is strongest for the first three principal components (see Supplement).

\begin{figure}
    \centering
    \includegraphics[width=0.95\linewidth]{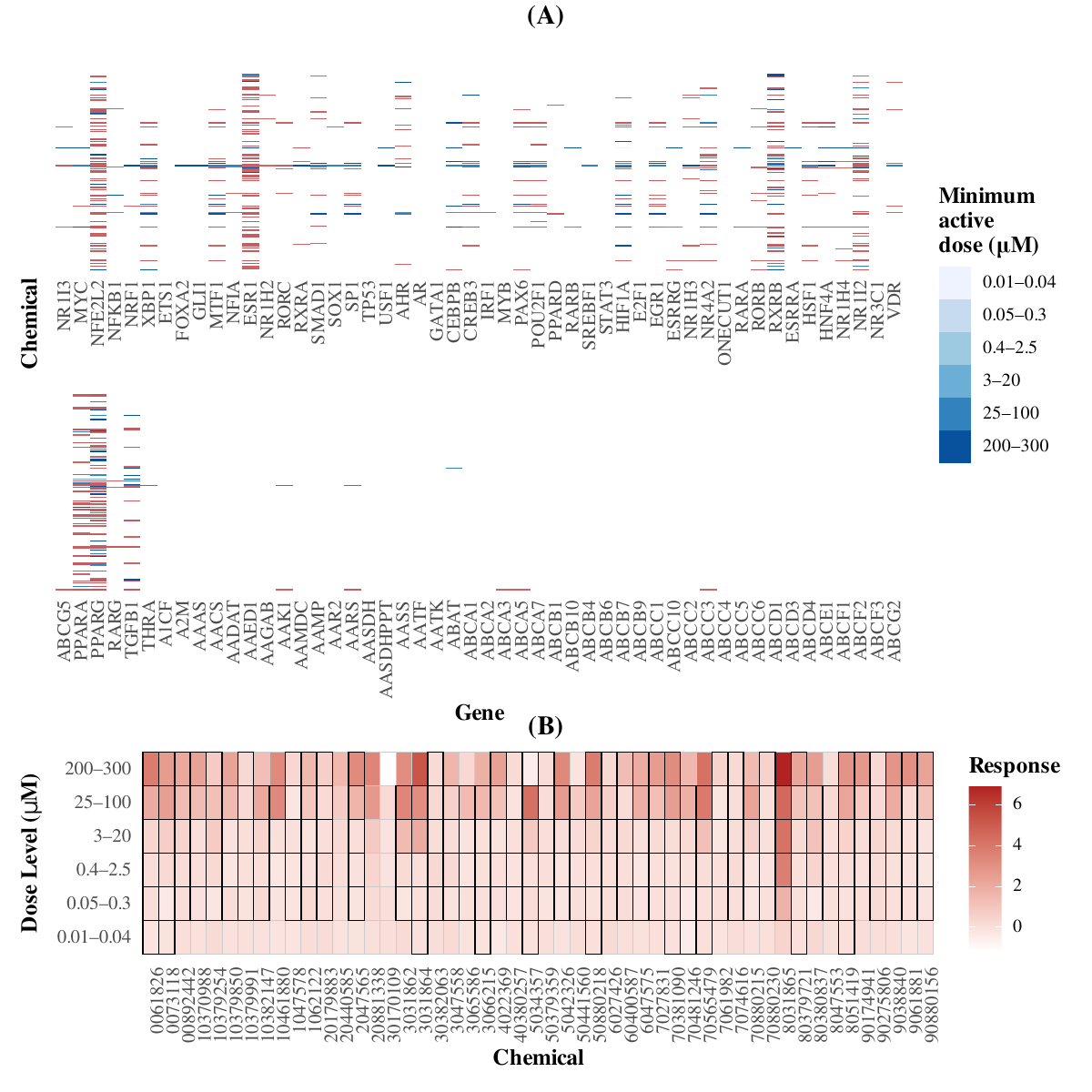}
    \caption{(A) Observed and newly predicted interactions: Posterior mean heatmap from full data fit showing minimum active dose to elicit a 75\% increase in activity over baseline. Blue shading indicates the posterior mean minimum active dose for newly predicted interactions (log-scale gradient). Red tiles mark pairs with observed activation in the data, and empty tiles denote pairs with neither observed nor predicted activation. (B) Predicted and observed PPARG assay expression responses across discretized dose levels for a subset of 50 chemicals (labeled via CASRN). Color indicates the response as log2-fold induction. Tiles with black outlines indicate observed  values; others are model predictions.}
    \label{fig:heatmap_post_combined}
\end{figure}

\begin{figure}
    \centering
\includegraphics[width=0.95\linewidth]{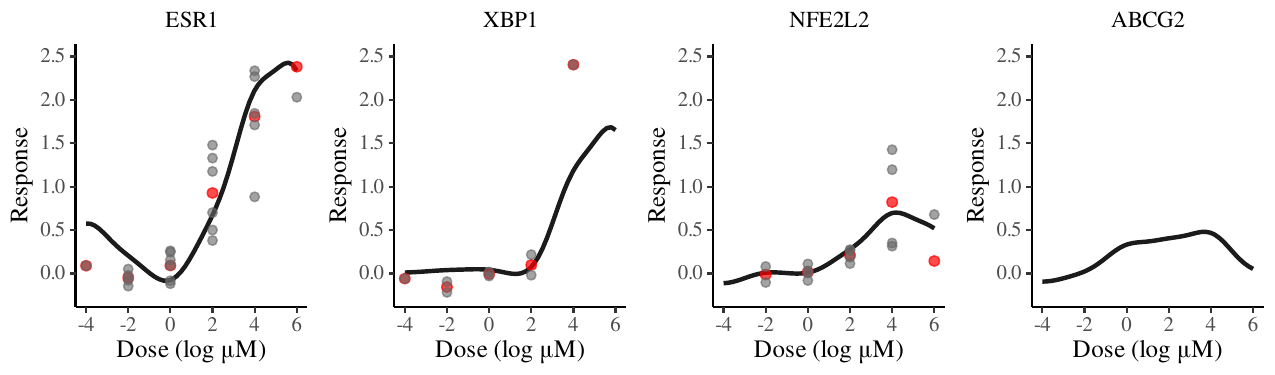}
    \caption{Posterior mean dose response curves for the chemical (Perfluorooctyl)ethanol (CASRN:678-39-7), a widely used barrier-forming agent, and a selection of genes with and without experimental data for the chemical. Response measured as log-2-fold induction. Observed data points are in grey with empirical mean response in red.}
    \label{fig:imputed_curves}
\end{figure}

A central objective of the proposed model is to prioritize PFAS chemicals for follow-up screening. In the observed data, 122 of the 193 chemicals show moderate activity for at least one gene in the subset, and 106 exhibit high activity—corresponding to 50\% and 75\% increases in assay expression over baseline, respectively. In the posterior dose–response network, the model identifies an additional four chemicals as moderately active and three as highly active for at least one gene assay. Figure \ref{fig:heatmap_post_combined} summarizes the high-activity findings: Panel (A) displays the observed and predicted minimum active dose required to elicit a 75\% expression increase, and Panel (B) provides a focused view of the PPARG assay, illustrating dose-specific responses across a range of PFAS, highlighting fine-scale differences in relative potency across PFAS species. Finally, the model imputes dose–response curves for chemical–assay pairs with partial or no observations (Fig. \ref{fig:imputed_curves}).

\section{Discussion}
\label{sec:dart_discussion}

We introduced a novel Bayesian framework (DART and its covariate-free version, DART-NC) for imputing missing dose–response curves from HTS studies. In a smaller subset where classical TCPL models could be fit, DART performed competitively, suggesting that the framework matches the strengths of existing approaches while extending to settings with far greater sparsity. Although the covariate-free model performed well on the larger, highly sparse HepG2–PFAS dataset, incorporating covariate information—chemical structural descriptors and gene ontology indicators—yielded modest improvements in predictive performance and provided additional interpretive structure. WAIC, LOO, and $k$-fold cross validation slightly favored DART over DART-NC, while CRPS suggested calibration was comparable across models. The strong baseline performance underscores the flexibility of the latent space formulation even without explicit prior information, though larger and more heterogeneous datasets may reveal greater advantages to covariate augmentation. 

Posterior inference highlighted biologically interpretable patterns, with detoxification, immune, and tissue genes estimated to be more responsive to PFAS exposure, and blood genes showing reduced sensitivity. While slightly more coefficients are significant on higher-index factors, our inferential focus remains on lower-index factors: under the MGP prior, higher‐index factors are shrunk a priori and typically carry less shared structure, tightening credible intervals so small shifts can appear significant while contributing little overall variance. Chemical effects were concentrated in a limited number of QSAR principal components, suggesting parsimony in structural drivers of activity. 

The DART model’s imputation and prioritization capacity has applied value. It identified additional chemicals likely to be active despite lacking direct experimental measurements, offering a principled tool for guiding follow-up screening. In some cases, the model predicts activity for genes without observed activation, driven by information borrowed across genes with shared ontologies. While such borrowing could overstate activity in sparsely represented pathways, it also highlights regions where functional priors exert the greatest influence and may warrant closer experimental investigation.

HTS data often exhibit idiosyncratic and context-dependent structure, and additional refinement is needed to fully accommodate the range of behaviors. More broadly, methodological development for functional matrix completion in environmental health remains limited. DART provides a flexible foundation for integrating information across heterogeneous sources of HTS data, is well suited to toxicological applications, and can readily generalize to other assays and associated covariates. This foundation can be developed further while already yielding practically useful predictions in high-throughput settings.

Posterior predictive diagnostics point to two main avenues for extension. First, the model tends to slightly over-smooth responses near baseline, resulting in too few strictly inactive curves and too many mild deviations. Introducing an explicit activity indicator or mixture component could better capture the spike at zero, and gene-level shrinkage within the hierarchical prior could help distinguish globally inactive genes. Second, coverage is slightly above nominal for the first four response quintiles but falls below nominal for the most extreme effects, indicating understated uncertainty in the tails. Heavier-tailed likelihoods or more flexible shrinkage in high-signal regions may improve tail calibration while preserving accuracy near the center of the distribution.

Although our application focuses on transcriptomic and reporter assays, the framework extends naturally to other widely collected HTS modalities, such as cytotoxicity and cell-health endpoints and enzyme-activity assays, which exhibit similar patterns of sparsity and shared structure. Moreover, many toxicogenomic studies now span multiple cellular contexts. Extending DART to jointly model multiple cell types—for example, combining the HepG2 PFAS data analyzed here with the recently released MCF7 transcriptomic dataset \citep{Koval}—would enable information sharing across cellular contexts.

\section{Acknowledgment}
This work was supported by the National Institutes of Health (R01ES035625), the NIH Summer Intern Program, the National Institute of Environmental Health Sciences (ZIA ES103368-02, T32ES007018, and P30ES010126). We thank Amy Herring and Filippo Ascolani for valuable feedback on an earlier draft.

\bigskip
\begin{center}
{\large\bf SUPPLEMENTARY MATERIAL}
\end{center}

\section{Simulation studies}
\label{sec:DART_SM_simulation}

To assess the model's ability to recover complex  dose-response surfaces, we prepared a simulation study under varying levels of missingness. Data are simulated from the DART model with a missingness level $\pi_\text{miss}$ ranging from 10\% to 75\%. Simulations use $N=100$ chemicals and $M = 100$ genes observed across $D=5$ distinct dose levels with three replicates per gene-chemical-dose triple. In each simulation, we mirror the highly challenging real world problem of inferring dose response curves for gene-chemical pairs completely lacking data: all replicates are masked for a random subset of $\pi_\text{miss}DNM$ dose-chemical-gene triples. Chemical and gene covariates are simulated with dimensions $P=5$, $Q=10$ respectively, and the dimension of the latent space is $K=10$. The complete specification of all hyperparameters is provided in Table \ref{tab:hyperparameters}.

\begin{figure}[htp]
    \centering
\includegraphics[width=0.99\linewidth]{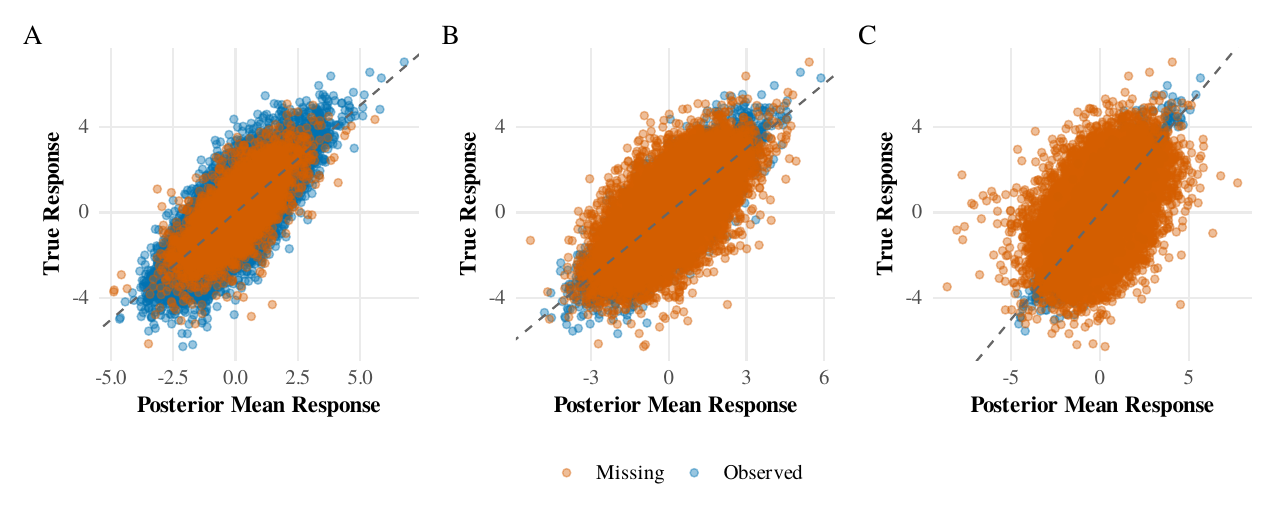}
    \caption{Comparison of posterior mean vs. true response in simulated data across three levels of missingness; color indicates observed data used for model fit vs missing data heldout for validation.
(A) 10\% missingness, (B) 50\% missingness, (C) 75\% missingness.}
    \label{fig:DART_SM_simulation_scatterplot}
\end{figure}

\begin{table}[ht]
\centering
\caption{Hyperparameter specifications for the DART model: simulations.}
\begin{tabular}{lll}
\hline
Variable & Interpretation & Prior / Value \\
\hline
$\boldsymbol{\mu}_{\text{gene}}$ & Mean vector for gene latent GPs (dim.~$D$) & $\mathbf{0}$ \\
$\sigma_\phi$ & SD of log-normal prior on gene-level scale matrix $\Phi$ & $0.25$ \\
$\delta_k$ & Multiplicative gamma process shrinkage ($\gamma_k$) & $\text{Gamma}(2,1), \; k=1,\dots,K$ \\
$\tau_0$ & Global shrinkage in $\Lambda$ & $\text{Gamma}(2,1)$ \\
$\sigma_\beta$ & SD for gene ontology coefficients & $0.5$ \\
$\sigma_\eta$ & SD for chemical latent factors $\eta$ & $1$ \\
$\sigma_W$ & SD for chemical covariate submodel & $1$ \\
$\sigma_\theta$ & SD for chemical QSAR coefficients $\Theta$ & $1$ \\
$l_c$ & Length scale for dose-level GP kernel & $1$ \\
$\alpha_D$ & Dose-dependent error model intercept & $N(\log(0.15), 0.5)$ \\
$\beta_D$ & Dose-dependent error model slope (log-SD vs.~dose index) & $N(0, 0.05)$ \\
$\tau_\gamma$ & Hierarchical variance prior on gene-specific error & $N_+(0, 0.5)$ \\
\hline
\end{tabular}
\label{tab:hyperparameters}
\end{table}

Figure \ref{fig:DART_SM_simulation_scatterplot} demonstrates strong in- and out-of-sample recovery with missingness levels up to 75\%. We observe only moderate declines in performance from low to high levels of missingness. In this pair holdout scheme, we have zero direct observations for each heldout pair, and hence prediction depends entirely on how well the gene and chemical latent factors are learned from other interactions involving the respective gene and chemical. Learning these parameters in turn depends on the overall size and connectivity of the observed graph; given a moderately large number of nodes, graph size is sufficient to permit inference even in high missingness scenarios. 

\clearpage

\section{PFAS-HepG2 application data details}\label{SM_app}

\subsection{Data preparation}\label{SM_app_data}
The PFAS-HepG2 data set consists of (1) the new UNC 2024 PFAS-HepG2 dataset (henceforth UNC24), and (2) all experiments involving PFAS and the HepG2 cell line available on the NIH's Integrated Chemical Environment on May 6, 2024 (henceforth ICE). 
The text below describes key data preparation steps required to standardize the results of these two sources for comparability.

\subsubsection{UNC24 PFAS Data}

The UNC24 PFAS mixtures dataset consists of experiments involving three different PFAS and 10,645 probe sets from the HepG2 immortalized liver cell line. The experiments expose the HepG2 cells to each of the three chemicals at concentrations of 2.5, 10, 25, and 35 $\mu M$. 
Each of the 10,645 probe sets is mapped to the expression of a single gene, while each gene is mapped to between 1 and 19 probe sets. Among the genes included, the vast majority (94\%) are mapped to a single probe set ID. In total, 10,247 distinct genes are represented. 
In addition to single chemical exposures, genes are exposed to an equimolar mixture of all three PFAS at the same concentrations; pairwise exposure experiments are not included. In the present work, we focus on the single chemical exposures and subset the data accordingly. All experiments are performed with six independent replicates and use a 24 hour exposure.

\subsubsection{ICE PFAS Data}

The raw ICE data include both gene-centric and non-gene-centric transcriptomic assay endpoints (e.g., enzyme inhibition, cytotoxicity). Gene-centric endpoints were mapped to genes using the \href{https://comptox.epa.gov/dashboard/assay-endpoints}{CompTox Chemicals Dashboard v2.4.1}, with relevant assays derived primarily from Attagene (ATG) and Tox21 (Toxicity Testing in the 21st Century). These endpoints span 77 individual genes across 73 unique endpoint–gene combinations: 65 correspond to a single gene, while the remainder map to between two and four genes.

Of the 77 genes, 28 are not present in the UNC24 data (AR, ARL, ESR1, ESR1L, ESR2, ESR2A, ESR2B, ESR2L, ESRRG, ETS1, FOS, GATA1, GLI1, LEF1, MYB, NR1H3, NR1I3, NR4A2, RARB, RARG, RORB, SOX1, TCF7L1, TFAP2A, TFAP2B, TFAP2D, THRAA, THRAL). Many of these genes are associated with estrogen and androgen receptor activity, making them of particular interest for assessing potential endocrine disruption by PFAS. Consistent with processing of the UNC24 data, all multi-gene assays were removed.

Finally, replicates are not explicitly identified in the ICE dataset. When replicate structure could be unambiguously inferred (i.e., when all doses were repeated consistently across replicates), we retained the data; otherwise, observations were dropped (333 of 15,303 total observations). 

Experimental protocols are highly heterogeneous with dose counts ranging from 1 to 30 (most with eight or fewer) and replication varying between 1 and 51 replicates (most with only two). Most experiments report log$_2$ fold induction measured at 24 hours, consistent with the UNC24 data.

\begin{figure}[htp!]
    \centering
    \includegraphics[width=0.75\linewidth]{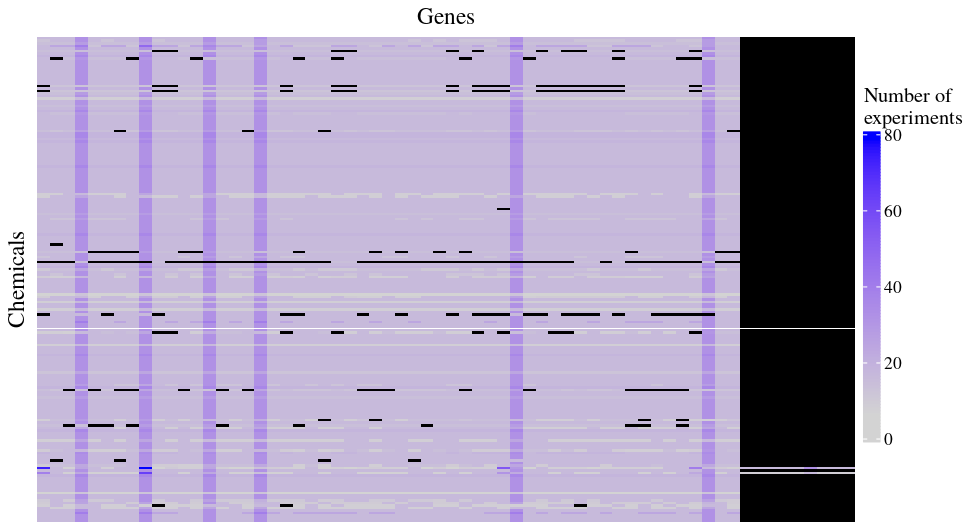}
    \caption{The cleaned ICE data consists of 193 PFAS chemicals (rows) and 64 single-gene assay endpoints (columns). Of these, 55 assay endpoints are tested on nearly all chemicals, with the remainder being tested on only three chemicals. Color indicates the number of experimental doses (including replicates) for each assay endpoint - chemical pair, with untested pairs indicated in black.}
    \label{fig:EDA_ICE_heatmap}
\end{figure}

\subsubsection{Dosage}
After discarding experiments involving multi-gene assays, chemical-mixtures, and percent activity responses, the combined data include experiments conducted at 40 unique dose levels from  0.1 to 300 $\mu$M. The number of dose levels per gene–chemical pair ranges from four to 21, with no chemical-gene pair observed at all 40 levels. 

Nearly all chemicals are tested at high concentrations (200–300$\mu$M) for at least some genes, though fewer than 1\% of genes are assayed at these extreme doses. Most experiments, comprising 75\% of chemical-gene pairs, are evaluated at only four concentrations. Although some marginal effects are observed at high doses, most activity appears to emerge at lower concentrations under 25 $\mu$M. 

\begin{table}[!h]
\centering
\begin{tabular}{cc}
\toprule
\textbf{Discretized log dose} & \textbf{Dose Range (µM)} \\
\midrule
–4 & 0.01 – 0.04 \\
–2 & 0.05 – 0.3 \\
0  & 0.4 – 2.5 \\
2  & 3 – 20 \\
4  & 25 – 100 \\
6  & 200 – 300 \\
\bottomrule
\end{tabular}
\caption{Dose ranges corresponding to discrete log-scaled bins used in the model.}
\label{tab:dose_bins}
\end{table}

We transform recorded doses by discretizing the natural logarithm of dose into six distinct dose-level bins; this procedure facilitates inference focused on the marginal increases at the (lower) doses of primary interest while reducing the dimension and sparsity of the experimental array. See Table \ref{tab:dose_bins} for the raw dose range for each bin. 

\subsubsection{Replicate effects}
Variability between replicates is generally low, with a small subset of experiments indicating instability or batch effects; this is generally consistent with the standardization expected with HTS involving genetically identical immortalized cell lines. 

\begin{figure}[!htp]
    \centering
\includegraphics[width=0.75\linewidth]{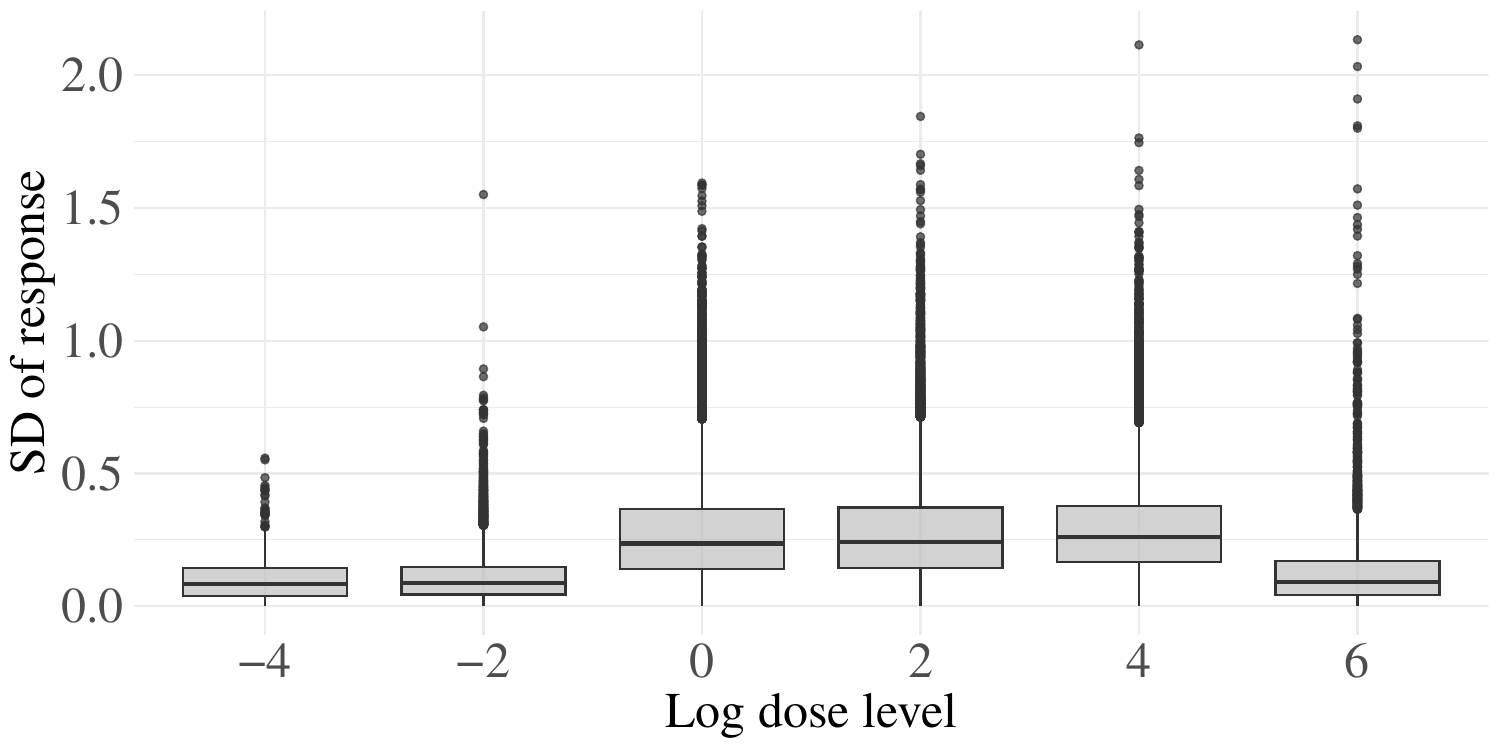}
    \caption{Between replicate standard deviation by dose level (natural logarithm).}
    \label{fig:dose_heterskedasticity}
\end{figure}
However, we see evidence of dose-dependent heteroskedasticity with higher variance at high dose levels; see Figure \ref{fig:dose_heterskedasticity}. 

\begin{figure}[!htp]
    \centering    \includegraphics[width=0.75\linewidth]{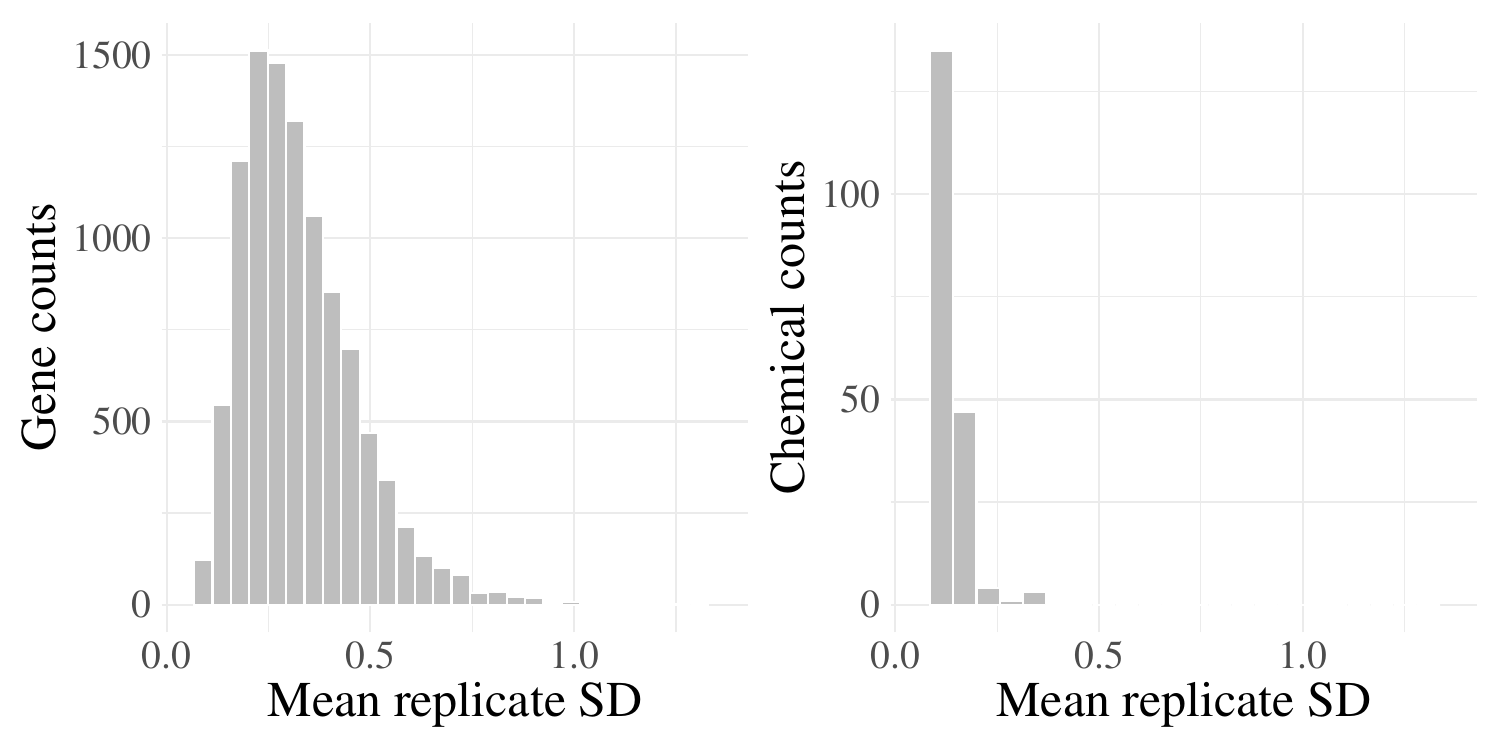}
    \caption{Distribution of replicate standard deviation across genes and chemicals.}
    \label{fig:gene_heteroskedasticy}
\end{figure}

Additionally we see substantial between-gene heterogeneity in baseline noise, with a much narrower spread in noise across chemicals; see Figure \ref{fig:gene_heteroskedasticy}.

\subsubsection{Response harmonization}

The response in HTS studies of chemical activity are typically reported as normalized log2-fold-induction, either with respect to a negative control reflecting baseline activity, or a positive control reflecting maximal expected induction: 

\begin{equation}
\label{eq:response_1}
\log_2\left(\frac{\text{expression}_{\text{sample}}}{\text{expression}_{\text{negative control}}}\right), 
\end{equation}

or 

\begin{equation}
\label{eq:response_2}
\log_2\left(\frac{\text{expression}_{\text{sample}} - \text{expression}_{\text{negative control}}}{\text{expression}_{\text{positive control}} - \text{expression}_{\text{negative control}}}\right).
\end{equation}

We use definition \ref{eq:response_1} and convert all observations accordingly. The UNC24 data are all reported as log2-expression, separately for both treatment and control; treatment and control values are merged to separately generate normalized log-2-fold induction for each experimental replicate. In the ICE data, the vast majority of responses are reported as``log2-fold-induction'' with the remaining reported as ``percent activity.'' ICE log-2-fold-induction is assumed to be with respect to a negative control as in definition \ref{eq:response_1} based on an implied percent activity in the range (0, 55) with mass strongly concentrated at 1, i.e. the null response in a study with negative control. Percent activity is assumed to be defined with respect to a positive control based on its observed range of (-111, 124), with mass strongly concentrated at 0, i.e., the null response in a study with a positive control. To ensure consistency in the response definition, ICE data reported as percent activity are discarded. 

\subsection{Gene set covariates} Gene set covariates are derived from the Hallmark Gene Sets. Of the 10,275 genes included in the combined dataset, 2,834 have at least one annotation among the 50 Hallmark Gene Sets. As covariates, the gene set data are highly sparse, with the median prevalence across the gene sets less than 1\%. To reduce the dimensionality of the covariate matrix and improve interpretability, we collapse the 50 gene sets into 12 pathways: tissue function, immune system, endocrine system, blood, cell regulation, digestion, metabolism, DNA, ATP, reproduction, and detoxification. The coarsened covariates consist of binary indicators equal to one if a gene is experimentally observed to express the function of the Hallmark Gene Sets associated with the given pathway and zero otherwise; pathway sets are non-disjoint, and genes in the current data occur in between zero and eight pathways. Given that genes may contribute to a pathway in yet undiscovered ways, we treat these data as presence-only.

\subsection{Chemical covariates}

Chemical covariates are derived from QSAR models of physiochemical properties through the \hyperlink{https://github.com/kmansouri/OPERA}{OPERA software}. We keep only point estimates of physicochemical properties, resulting in a collection of 32 real-valued chemical covariates. The selected covariates capture a broad range of chemical, physical, and biological properties. Structural descriptors include atom counts, ring structures, hybridization, and molecular weight. Physicochemical properties span melting and boiling points, surface area, and polarizability. Partitioning and lipophilicity are represented by logP, logD, and KOA values; volatility and solubility by vapor pressure and water solubility; and environmental fate by predicted half-life and retention time. Additional covariates model ionization and pKa, while ADME-related variables capture predicted clearance, permeability, and plasma binding. OPERA also provides applicability domain flags, index scores, and confidence estimates, but only point estimates were retained for this analysis. The extracted features are described in Tables~\ref{tab:SM_qsar_physchem}--\ref{tab:SM_qsar_adme}. 

\begin{table}[ht]
\centering
\small
\caption{Physicochemical descriptors including molecular structure and ionization.}
\label{tab:SM_qsar_physchem}
\begin{tabular}{ll}
\toprule
\textbf{Variable} & \textbf{Description} \\
\midrule
MolWeight & Molecular weight \\
nbAtoms & Number of atoms \\
nbHeavyAtoms & Number of heavy atoms \\
nbC & Number of carbon atoms \\
nbO & Number of oxygen atoms \\
nbN & Number of nitrogen atoms \\
nbAromAtom & Number of aromatic atoms \\
nbRing & Number of rings \\
nbHeteroRing & Number of heteroatom-containing rings \\
Sp3Sp2HybRatio & Ratio of sp3 to sp2 hybridized atoms \\
nbRotBd & Number of rotatable bonds \\
nbHBdAcc & Number of hydrogen bond acceptors \\
ndHBdDon & Number of hydrogen bond donors \\
nbLipinskiFailures & Lipinski rule violations \\
TopoPolSurfAir & Topological polar surface area \\
MolarRefract & Molar refractivity \\
CombDipolPolariz & Combined dipole polarizability \\
ionization & Ionization state \\
\bottomrule
\end{tabular}
\end{table}

\begin{table}[ht]
\centering
\small
\caption{Descriptors of partitioning behavior and environmental fate.}
\label{tab:SM_qsar_partition_env}
\begin{tabular}{ll}
\toprule
\textbf{Variable} & \textbf{Description} \\
\midrule
LogP\_pred & Predicted logP (octanol–water partition coefficient) \\
LogD55\_pred & Predicted logD at pH 5.5 \\
LogD74\_pred & Predicted logD at pH 7.4 \\
LogKOA\_pred & Predicted log KOA (octanol–air partition coefficient) \\
LogVP\_pred & Predicted log vapor pressure \\
LogWS\_pred & Predicted log water solubility \\
LogHL\_pred & Predicted log half-life \\
RT\_pred & Predicted retention time \\
\bottomrule
\end{tabular}
\end{table}

\begin{table}[ht]
\centering
\small
\caption{Biological descriptors related to absorption, distribution, metabolism, and excretion (ADME).}
\label{tab:SM_qsar_adme}
\begin{tabular}{ll}
\toprule
\textbf{Variable} & \textbf{Description} \\
\midrule
FUB\_pred & Predicted unbound fraction in plasma \\
Clint\_pred & Predicted intrinsic clearance \\
CACO2\_pred & Predicted Caco-2 cell permeability \\
\bottomrule
\end{tabular}
\end{table}

\begin{figure}[ht]
    \centering
    \includegraphics[width=0.95\linewidth]{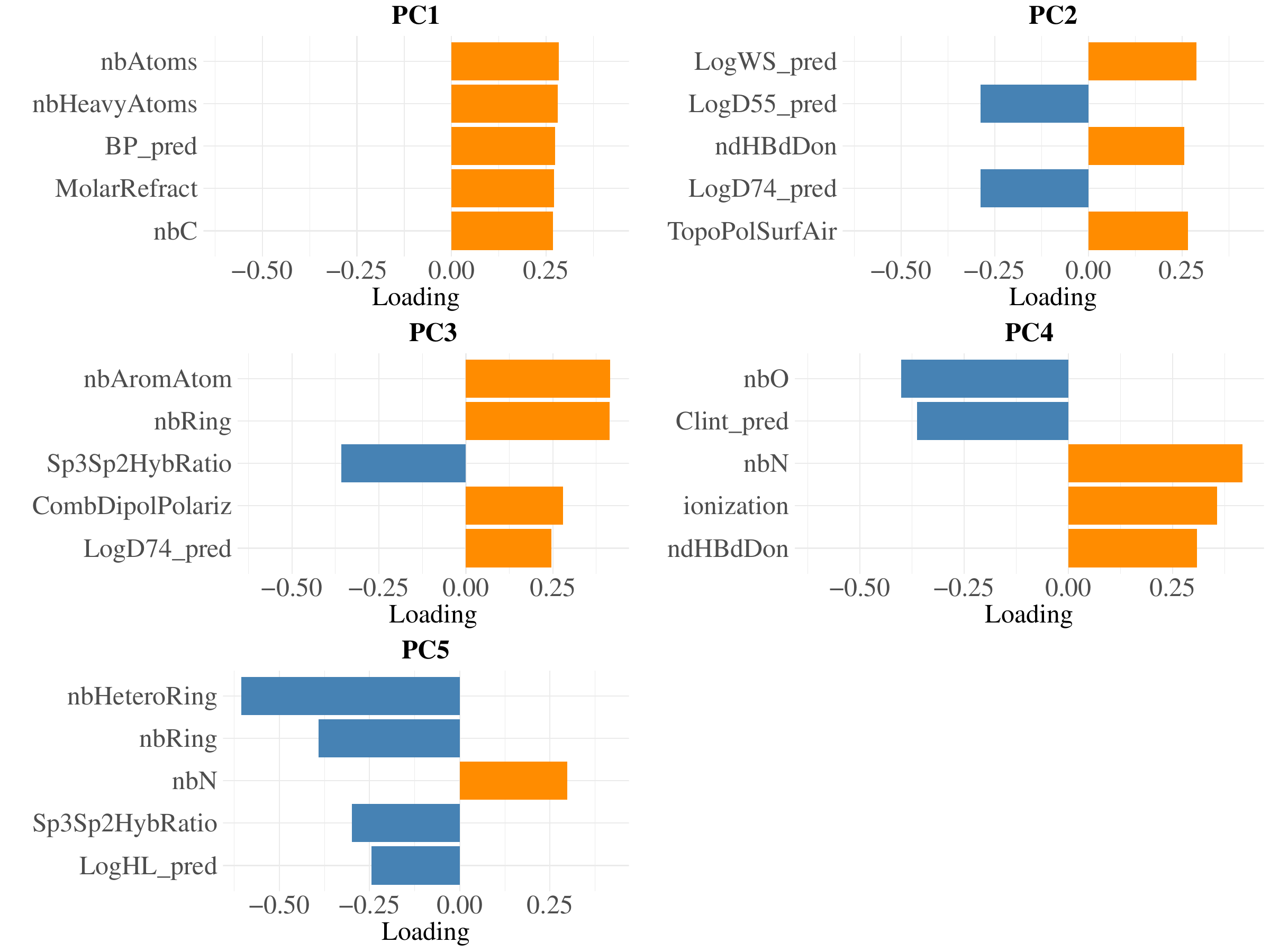}
    \caption{The top QSAR covariate loadings for the five leading principal components.}
    \label{fig:qsar_loadings}
\end{figure}

\begin{figure}[ht]
    \centering
    \includegraphics[width=0.95\linewidth]{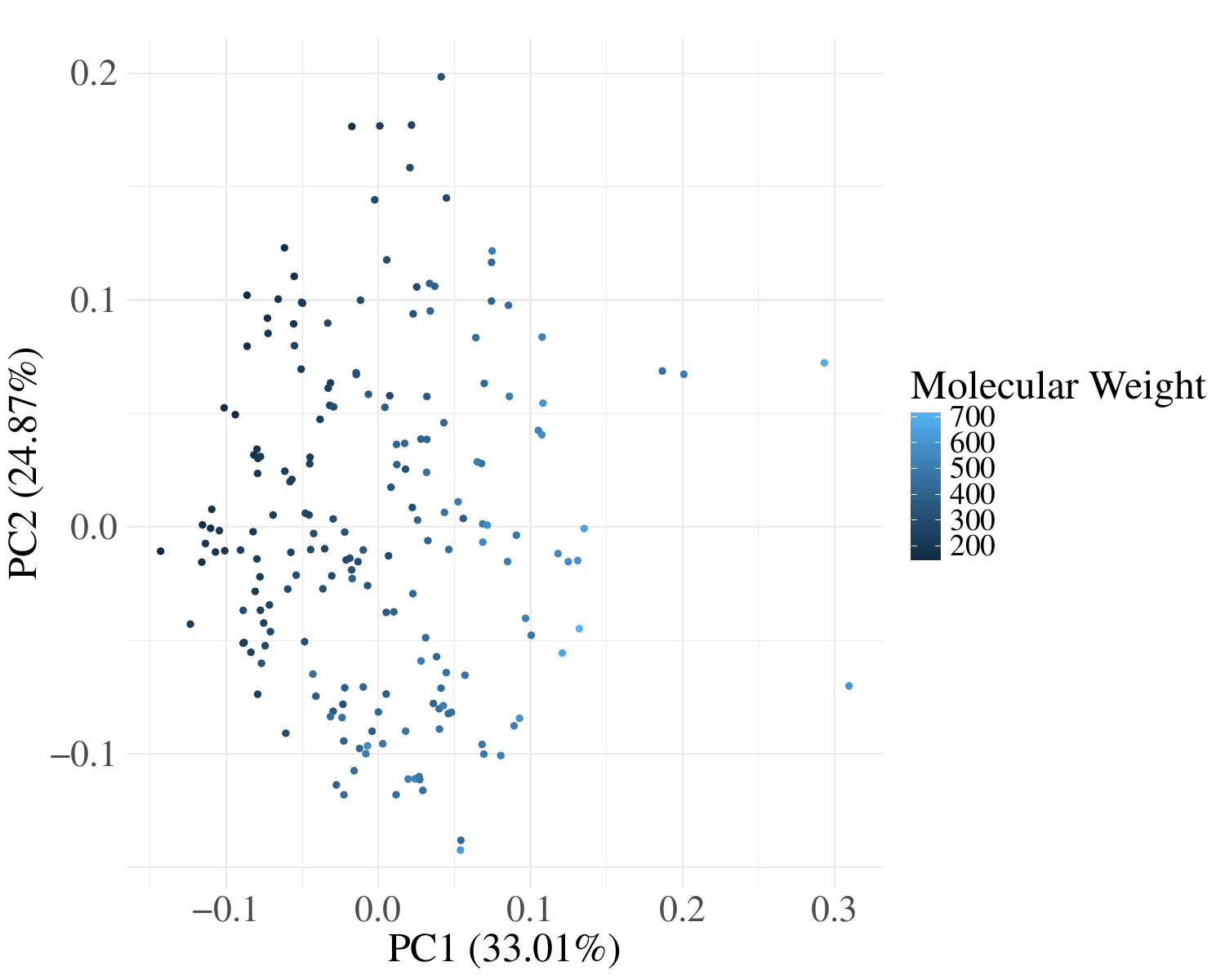}
    \caption{The leading two QSAR principal components versus molecular weight.}
    \label{fig:qsar_scatterplot}
\end{figure}

We reduce these to the first five principal components, explaining 76.7\% of the variance. We remove two chemicals not present in OPERA from the data, together consisting of fewer than 0.1\% of experimental observations. The top QSAR covariate loadings for the leading principal components are illustrated in Figure \ref{fig:qsar_loadings}. Chemicals are plotted with respect to the first two principal components in Figure \ref{fig:qsar_scatterplot}.  

\clearpage

\section{Comparison with classical TCPL models}

\subsection{Specification of classical benchmark models}

Additional details of the benchmark models utilized from the \texttt{tcplfit2} package are provided below. Each model is fit, using the \texttt{tcplfit2} software, to a subset of chemicals and genes each containing at least four distinct doses and run until convergence or failure; fitted model parameters are then used to generate out of sample predictions (not permitted in the software). 

Classical dose-response models estimate key parameters that describe biological activity, such as AC$_{50}$ (the concentration at which the half maximum effect is elicited), E$_{\text{max}}$ (the maximum response), and additional parameters from mechanistic toxicokinetic models. These models, including the popular Hill, Exponential 5, and Power functions, typically fit each curve independently and assume a parametric form with a small number of parameters, and generally cannot capture non-monotonic responses \citep{judson2010invitro, thomas2019blueprint}. Although classical dose-response models are straightforward to implement and interpret, they often fall short in high-throughput screening (HTS) contexts, where data are sparse, noisy, and frequently deviate from monotonic response patterns. 

The Hill function was originally developed in pharmacology and models a sigmoidal relationship between dose concentration and response \citep{hill1910}. Today, it is widely used in biology, pharmacology, and toxicology. The Hill function is defined as

$$f(x) = \frac{tp}{1 + \left(\frac{AC_{50}}{x}\right)^p}$$
where $x$ is the dose, $tp$ is the top or theoretical maximum response, $AC_{50}$ is the concentration at half-maximal response, and $p$ is the Hill power coefficient controlling the steepness of the curve, with large values of $p$ yielding a sharp step centered at the $AC_{50}.$

The Exponential 5 (exp5) function was developed specifically for HTS screening applications within the ToxCast program; it is a flexible model capable of capturing monotonic increasing or decreasing trends. The exp5 model is defined as:

$$f(x) = tp*(1-2^{-(x/AC_{50})^p})$$
where $tp$ and $AC_{50}$ have the same interpretation as in the Hill function, and the power parameter $p$ controls the exponential rise from the null to the maximal response. In general, exp5 does not plateau as sharply as the Hill function outside of extreme values of $p$. 

The Power function models unbounded, monotonic trends with accelerating or decelerating behavior and is expressed as
$f(x) = a x^p$,
where $a$ is a scaling parameter and the power term $p$ controls the curvature of the response, without reference to any inflection point like the $AC_{50}$. 

The applicability of these models varies depending on the biological system of interest and the shape of the dose–response relationship. In general, the Hill model is preferred for an expected sigmoidal response with symmetry around the $AC_{50}$, particularly when there is interest in estimating the $AC_{50}$ itself. The Exp5 function is preferred for asymmetric, sigmoidal curves, with flat response prior to a threshold, and steep activation above it. As with the Hill model, the Exp5 response is bounded but permits sharper transitions. The power function is preferred for monotonic and unbounded responses, and is useful for fitting data with observations at only a few doses.

\subsection{DART specification}
DART hyperparameters are the same as those used in the simulations, with the exception of those listed in Table \ref{tab:hyperparameters_pfas}.

\begin{table}[ht]
\centering
\caption{Hyperparameter specifications for the DART model.}
\begin{tabular}{lll}
\hline
Variable & Interpretation & Prior / Value \\
\hline
$\sigma_\phi$ & SD of log-normal prior on gene-level scale matrix $\Phi$ & $0.15$ \\
$\delta_k$ & Multiplicative gamma process shrinkage ($\gamma_k$) & $\text{Gamma}(5,2), \; k=1$ \\
$\delta_k$ & Multiplicative gamma process shrinkage ($\gamma_k$) & $\text{Gamma}(5,2.5), \; k = 2, \dots,K$ \\
$\tau_0$ & Global shrinkage in $\Lambda$ & $\text{Gamma}(3,2)$ \\
\hline
\end{tabular}
\label{tab:hyperparameters_pfas}
\end{table}

\subsection{Performance details by fold}

Here, we analyze the performance of the proposed DART model and TCPL competitor models by fold, where fold $k$ holds out the $k$th lowest concentration tested for each chemical-gene pair if sufficient doses are present to allow fitting of the TCPL models and no doses are held out otherwise. If a given chemical-gene pair has fewer than five observed concentration levels, no doses will be heldout on some fold and all data for the corresponding chemical-gene pair are ``in-sample'' and excluded from performance metrics for that fold. We see that DART, Exp5, and Hill models generally perform similarly, and have the weakest performance when intermediate doses are heldout. This is to be expected: at low concentrations, most dose-response curves flatten towards the baseline, while at the high end, curves often saturate to a peak effect, whereas intermediate doses typically correspond to the regions of steep slope and determine the overall curve shape. The power model performs worse overall and does especially poorly when high doses are omitted, resulting from the fact that its functional form does not explicitly include an upper asymptote parameter.

\begin{figure}
    \centering
    \includegraphics[width=\linewidth]{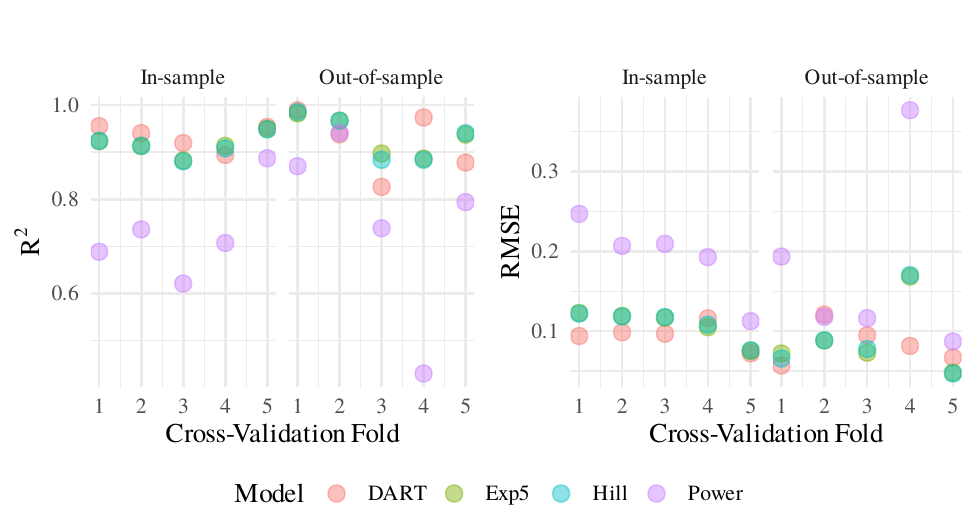}
    \caption{By-fold comparison of model performance in a small, dense subset of the data consisting of ten genes and ten chemicals. Outlier poor performance for the Power model in fold 5 is excluded from performance metrics for ease of visualization. }
    \label{fig:tcpl_by_fold}
\end{figure}

\clearpage

\section{PFAS-HepG2 Application (Full Dataset)}

 \subsubsection{Additional results: Chemical QSAR coefficients}
\begin{figure}[!htp]
    \centering
    \includegraphics[width=0.95\linewidth]{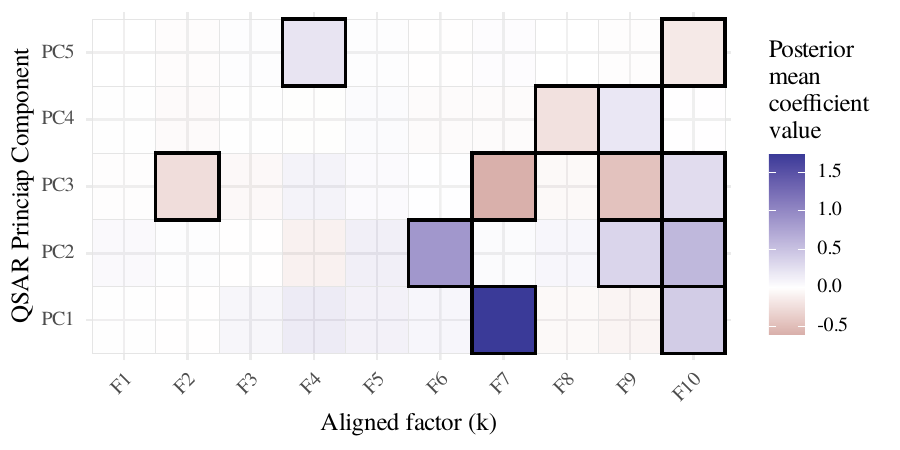}
    \caption{Chemical QSAR covariate effects on aligned latent factors: color indicates posterior mean $\theta_{pk}$; outlined cells have 95\% credible intervals excluding 0.}
    \label{fig:qsar_theta}
\end{figure}

 Figure \ref{fig:qsar_theta} shows the posterior mean values of the QSAR submodel coefficients, $\theta_{lk}$. Covariate effects are concentrated in the first three principal components, with smaller magnitude effects associated with pricipal components four and five. 

\clearpage

\subsection{Model Diagnostics}

We assessed MCMC convergence using standard summaries of the log posterior density (total log joint probability of parameters and data). Trace plots show stable mixing. The estimated potential scale reduction factor was $\hat{R} = 1.01$, and the effective sample sizes were 2769 (bulk) and 4285 (tail), indicating adequate exploration of the posterior distribution.

 \begin{figure}[!htp]
     \centering
     \includegraphics[width=0.75\linewidth]{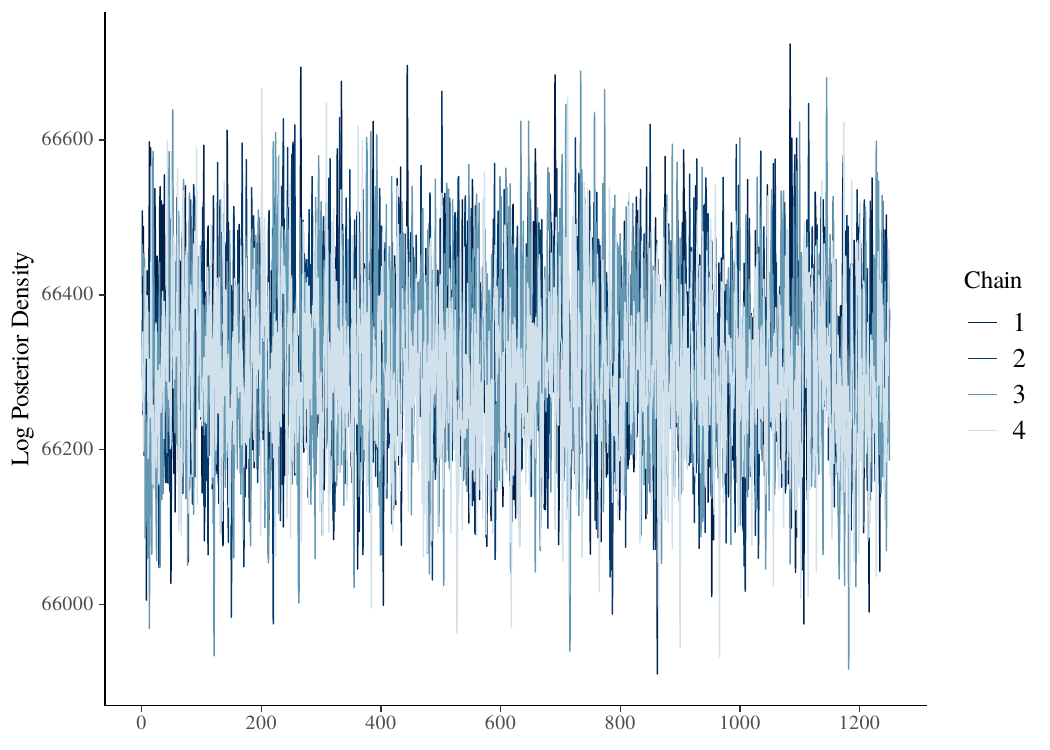}
     \caption{Trace of log posterior density, total log join probability of parameters and data. Rhat is 1.01, ESS (effective sample size) bulk 2769, ESS tail 4285.}
     \label{fig:lp_trace}
 \end{figure}

\begin{figure}[!htp]
     \centering
\includegraphics[width=0.75\linewidth]{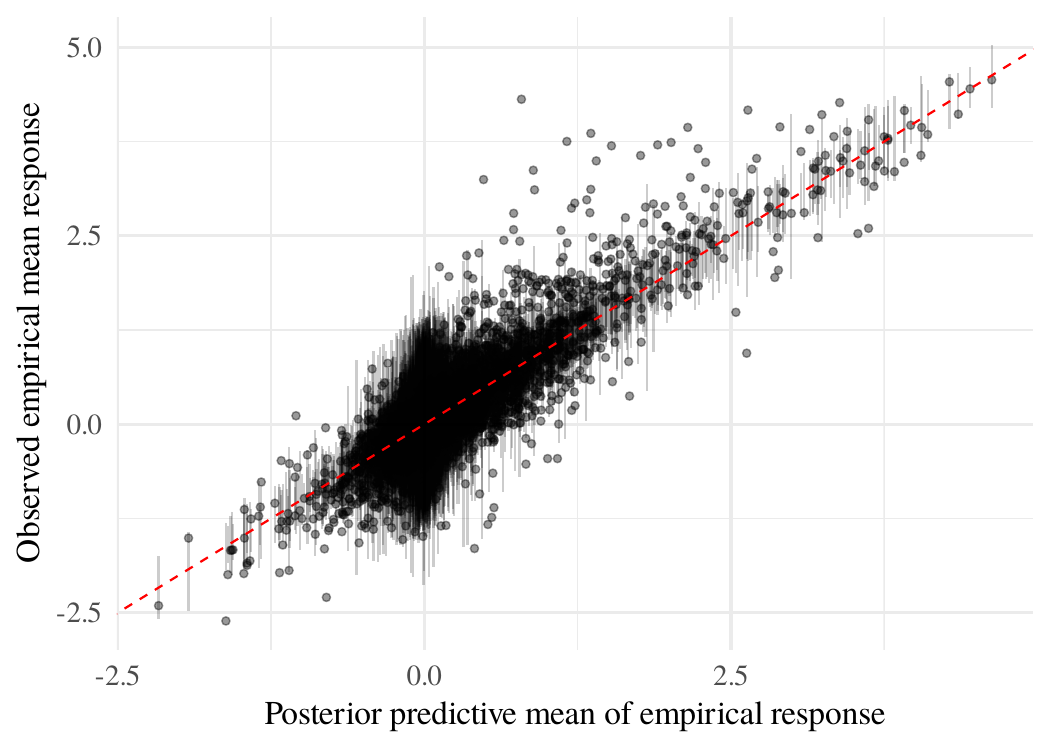}
     \caption{Observed versus posterior mean predicted responses with 95\% posterior predictive credible intervals.}
     \label{fig:ppc_scatterplot}
 \end{figure}

Posterior predictive diagnostics indicate that the model captures the overall magnitude and variability of responses well. Figure \ref{fig:ppc_scatterplot} shows observed empirical mean versus posterior predictive mean responses, with 95\% credible intervals. The points lie close to the 45 degree line, suggesting that the model is well calibrated on average. Wider posterior credible intervals near zero reflect greater model uncertainty about whether small deviations represent true chemical activity or background noise; intervals are narrow for more strongly active chemicals where the signal is clearer.  95\% posterior predictive coverage with respect to empirical means is close to nominal at 92.5\%.

\begin{figure}[!htp]
    \centering
    \includegraphics[width=0.65\linewidth]{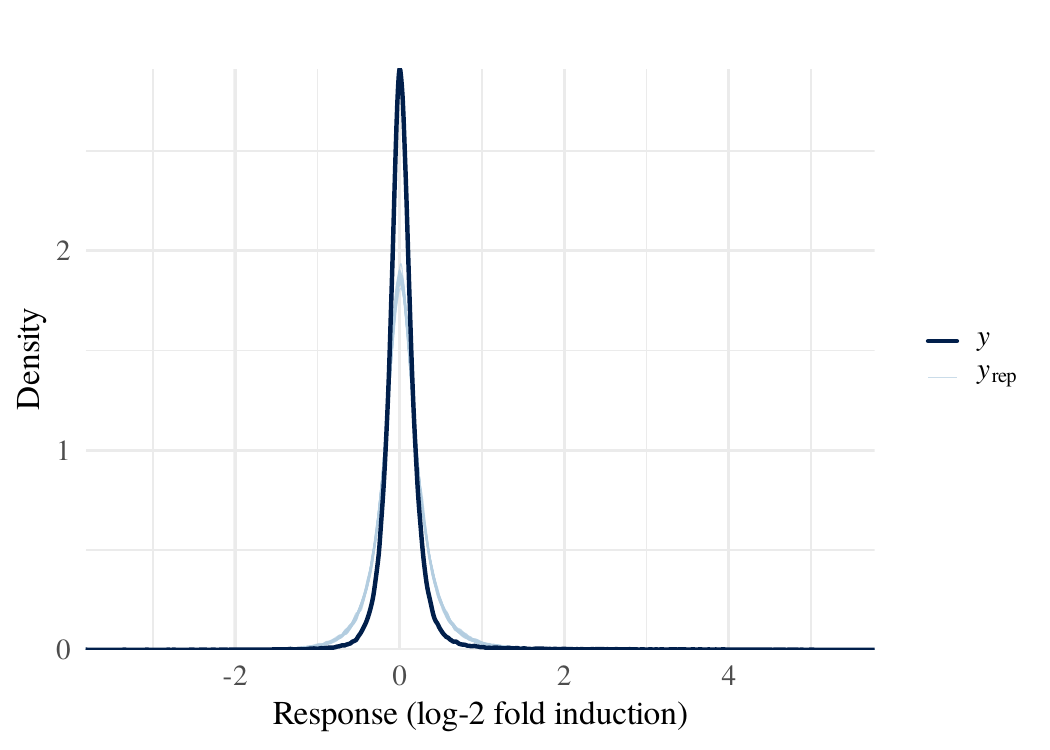}
    \caption{Posterior predictive vs observed distribution of log-2 fold induction. }
    \label{fig:ppc_density}
\end{figure}

Having shown that pointwise predictions align well, we next assess whether the predictive distribution reproduces the overall shape of the observed responses. Figure \ref{fig:ppc_density} shows that the model slightly underestimates inactive responses and overpredicts mild activity, reflecting smoothing near baseline but good fit for strong signals. The posterior predictive density underestimates the sharp spike of non-responsive genes at baseline while overpredicting mild deviations ($|\text{log}_2 \text{fold}| \approx 0.25–1$), reflecting a tendency to smooth over exact inactivity.

Finally, we investigate calibration by response magnitude to further explore behavior observed above. Coverage slightly exceeds nominal levels for weak and moderate responses. In contrast, coverage drops below nominal for strongly induced responses, indicating that uncertainty is somewhat underestimated in the tails. 

\begin{figure}[!htp]
    \centering
\includegraphics[width=0.65\linewidth]{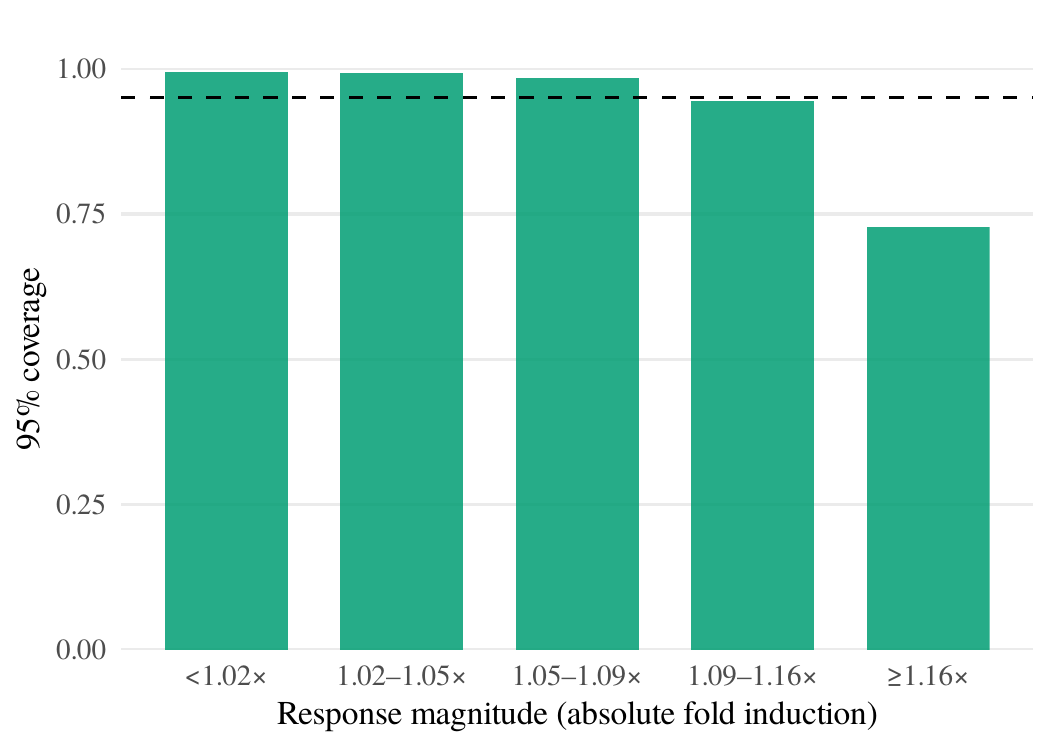}
    \caption{Posterior predictive 95\% coverage by response magnitude. Bins correspond to quintiles of the absolute fold induction relative to baseline response. }
    \label{fig:placeholder}
\end{figure}

\clearpage 

\subsection{Screening recommendations}
Table \ref{tab:screening} presents a subset of chemicals that showed no observed activity in the data at one or more of the three thresholds (25 \%, 50 \%, and 75 \% activity above baseline) but were predicted as active at the corresponding threshold(s) by the DART model. Chemicals are ranked by their median SEEM3 Consensus total population exposure hazard, estimated using the \texttt{ctxR} package \citep{ctxR}. Compounds with comparatively high exposure hazards (greater than $10^{-6}$ mg kg$^{-1}$ day$^{-1}$) should be considered priority candidates for further evaluation.


\centering
\begin{table}[!h]
\centering
\caption{Prioritized chemicals with SEEM3 exposure estimates and new predicted gene activity}
\centering
\resizebox{\ifdim\width>\linewidth\linewidth\else\width\fi}{!}{
\begin{tabular}[t]{lllccc}
\toprule
Compound & DTXSID & \makecell[c]{Gene with\\ Max Response} & 
\makecell[c]{SEEM3 Exposure\\(mg/kg/day)} & 
\makecell[c]{Predicted\\Active Genes} & 
\makecell[c]{Predicted\\Highly Active Genes}\\
\midrule
\cellcolor{gray!10}{Sevoflurane} & \cellcolor{gray!10}{8046614} & \cellcolor{gray!10}{ABCG2} & \cellcolor{gray!10}{5.4e-06} & \cellcolor{gray!10}{3} & \cellcolor{gray!10}{0}\\
Perfluoropropyl trifluorovinyl ether & 0061826 & ABCB10 & 3.6e-06 & 3 & 0\\
\cellcolor{gray!10}{2-Amino-2H-perfluoropropane} & \cellcolor{gray!10}{70481246} & \cellcolor{gray!10}{ABCB10} & \cellcolor{gray!10}{2.3e-06} & \cellcolor{gray!10}{3} & \cellcolor{gray!10}{0}\\
(Heptafluorobutanoyl)pivaloylmethane & 3066215 & PPARG & 1.3e-06 & 9 & 1\\
\cellcolor{gray!10}{6:1 Fluorotelomer alcohol} & \cellcolor{gray!10}{00190950} & \cellcolor{gray!10}{ESR1} & \cellcolor{gray!10}{1.2e-06} & \cellcolor{gray!10}{5} & \cellcolor{gray!10}{0}\\
\addlinespace
Difluoromethyl 1H,1H-perfluoropropyl ether & 0074059 & ABCB10 & 1.1e-06 & 3 & 0\\
\cellcolor{gray!10}{Allyl perfluoroisopropyl ether} & \cellcolor{gray!10}{10370988} & \cellcolor{gray!10}{ABCG2} & \cellcolor{gray!10}{1.1e-06} & \cellcolor{gray!10}{3} & \cellcolor{gray!10}{0}\\
Perfluorobutyraldehyde & 10190946 & ABCG2 & 9.0e-07 & 5 & 0\\
\cellcolor{gray!10}{Pentafluoropropionamide} & \cellcolor{gray!10}{0059871} & \cellcolor{gray!10}{ABCG2} & \cellcolor{gray!10}{9.0e-07} & \cellcolor{gray!10}{4} & \cellcolor{gray!10}{0}\\
Methyl perfluorobutanoate & 4059881 & ABCG2 & 9.0e-07 & 3 & 0\\
\addlinespace
\cellcolor{gray!10}{1-Propenylperfluoropropane} & \cellcolor{gray!10}{70379270} & \cellcolor{gray!10}{ABCB10} & \cellcolor{gray!10}{8.0e-07} & \cellcolor{gray!10}{3} & \cellcolor{gray!10}{0}\\
Methyl perfluoroethyl ketone & 90285748 & ABCB10 & 8.0e-07 & 3 & 0\\
\cellcolor{gray!10}{2-(Perfluorooctyl)ethanthiol} & \cellcolor{gray!10}{20337446} & \cellcolor{gray!10}{ABCB10} & \cellcolor{gray!10}{7.0e-07} & \cellcolor{gray!10}{3} & \cellcolor{gray!10}{0}\\
1H,1H,2H-Perfluorocyclopentane & 50880218 & ABCB10 & 7.0e-07 & 3 & 0\\
\cellcolor{gray!10}{1H,1H-Perfluoroheptylamine} & \cellcolor{gray!10}{10379835} & \cellcolor{gray!10}{ABCB10} & \cellcolor{gray!10}{6.0e-07} & \cellcolor{gray!10}{4} & \cellcolor{gray!10}{0}\\
\addlinespace
1-Bromoheptafluoropropane & 3059971 & ABCB10 & 5.0e-07 & 3 & 0\\
\cellcolor{gray!10}{Hexafluoroamylene glycol} & \cellcolor{gray!10}{3059927} & \cellcolor{gray!10}{ABCB10} & \cellcolor{gray!10}{5.0e-07} & \cellcolor{gray!10}{3} & \cellcolor{gray!10}{0}\\
Perfluorocyclohexanecarbonyl fluoride & 80379781 & ABCB10 & 4.0e-07 & 3 & 0\\
\cellcolor{gray!10}{3-(Perfluorohexyl)-1,2-epoxypropane} & \cellcolor{gray!10}{30880413} & \cellcolor{gray!10}{PPARG} & \cellcolor{gray!10}{3.0e-07} & \cellcolor{gray!10}{8} & \cellcolor{gray!10}{1}\\
Perfluorobutanesulfonyl fluoride & 20861913 & ABCB10 & 3.0e-07 & 3 & 0\\
\addlinespace
\cellcolor{gray!10}{Trichloro((perfluorohexyl)ethyl)silane} & \cellcolor{gray!10}{50229163} & \cellcolor{gray!10}{ABCB10} & \cellcolor{gray!10}{3.0e-07} & \cellcolor{gray!10}{3} & \cellcolor{gray!10}{0}\\
2-(Perfluorohexyl)ethanethiol & 20379947 & ESR1 & 2.0e-07 & 8 & 1\\
\cellcolor{gray!10}{2-(Perfluorohexyl)ethylphosphonic acid} & \cellcolor{gray!10}{20179883} & \cellcolor{gray!10}{ABCG2} & \cellcolor{gray!10}{2.0e-07} & \cellcolor{gray!10}{3} & \cellcolor{gray!10}{0}\\
Dichloromethyl((perfluorohexyl)ethyl)silane & 00223797 & PPARG & 1.0e-07 & 5 & 0\\
\cellcolor{gray!10}{3,3,4,4,5,5,6,6,6-Nonafluorohexene} & \cellcolor{gray!10}{6047575} & \cellcolor{gray!10}{ABCB10} & \cellcolor{gray!10}{1.0e-07} & \cellcolor{gray!10}{3} & \cellcolor{gray!10}{0}\\
\addlinespace
Ethyl perfluorobutyl ether & 0073118 & ABCB10 & 1.0e-07 & 3 & 0\\
\bottomrule
\end{tabular}}
\label{tab:screening}
\end{table}

\bibliography{paper-ref}
\bibliographystyle{apalike}

\end{document}